\documentclass[12pt,showpacs,pra,superscriptaddress]{revtex4-1}
\usepackage{amssymb}
\usepackage{amsfonts}
\usepackage{amsmath}

\usepackage{bm}
\usepackage{epsfig}
\usepackage{epsf}
\usepackage[]{graphicx}
\usepackage[sort&compress]{natbib}
\usepackage{color}
\usepackage{xspace}

\renewcommand {\d} {{\rm d}}
\renewcommand {\i} {{\rm i}}

\newcommand {\ee}  {{\rm e}}

\newcommand {\E}  {{\varepsilon}}
\newcommand {\om}  {{\omega}}
\newcommand {\Om}  {{\Omega}}

\newcommand {\omp}  {{\omega^{\prime}}}

\newcommand {\bfbeta} {\boldsymbol{\beta}}
\newcommand {\bnabla} {\boldsymbol{ \nabla}}
\newcommand {\btheta} {\boldsymbol{ \theta}}
\newcommand {\brho} {\boldsymbol{\rho}}
\newcommand {\bDelta} {\boldsymbol{\Delta}}

\newcommand {\bfa} {{\bf a}}

\newcommand {\bfe} {{\bf e}}

\newcommand {\bfn} {{\bf n}}
\newcommand {\bfp} {{\bf p}}
\newcommand {\bfr} {{\bf r}}
\newcommand {\bfv} {{\bf v}}

\newcommand {\bfE} {{\bf E}}
\newcommand {\bfF} {{\bf F}}

\newcommand {\bfP} {{\bf P}}
\newcommand {\bfR} {{\bf R}}
\newcommand {\bfT} {{\bf T}}

\newcommand {\calA}  {{\cal A}}
\newcommand {\calF}  {{\cal F}}

\newcommand {\Zp} {Z_{\rm p}}

\newcommand {\dUmax} {{U_{\max}^{\prime}}}

\newcommand {\Uat} {U_{\rm at}}
\newcommand {\aTF} {a_{\rm TF}}

\newcommand {\Ld} {L_{\rm d}}
\newcommand {\Lp} {L_{\rm p}}

\newcommand{\MBNExplorer}{\textsc{MBN Explorer}\xspace}

\begin{document}

\title{Simulation of Ultra-Relativistic Electrons and Positrons Channeling 
in Crystals with \MBNExplorer}

\author{Gennady B. Sushko}
\affiliation{Frankfurt Institute for Advanced Studies, Goethe University,
Ruth-Moufang-Str. 1,
 60438 Frankfurt am Main, Germany}
\affiliation{Virtual Institute on Nano Films (VINF), 
All\'{e}e des Noisetiers, 2 bte 30 Angleur, Belgium}
\author{Victor G. Bezchastnov
\footnote{On leave from 
Department of Theoretical Astrophysics, 
Ioffe Physical-Technical Institute, Politechnicheskaya Str. 26, 
194021 St. Petersburg, Russia
and from St. Petersburg State Polytechnical University,
Politechnicheskaya  29, 195251,  St. Petersburg, Russia}}
\affiliation{Frankfurt Institute for Advanced Studies, Goethe University,
Ruth-Moufang-Str. 1, 60438 Frankfurt am Main, Germany}
\author{Ilia A. Solov'yov\footnote{On leave from 
Ioffe Physical-Technical Institute,
Politechnicheskaya  26, 194021 St. Petersburg, Russia}
}
\affiliation{Virtual Institute on Nano Films (VINF), 
All\'{e}e des Noisetiers, 2 bte 30 Angleur, Belgium}
\affiliation{Beckman Institute for Advance Science and Technology, 
University of Illinois\\
 at Urbana-Champaign, 405 N. Mathews Ave, Urbana Illinois 61801, USA}
\author{Andrei V. Korol\footnote{E-mail: korol@fias.uni-frankfurt.de}}
\affiliation{Frankfurt Institute for Advanced Studies, Goethe University,
Ruth-Moufang-Str. 1, 60438 Frankfurt am Main, Germany}
\affiliation{St. Petersburg State Maritime University, 
Leninsky ave. 101, 198262 St. Petersburg, Russia}
\author{Walter Greiner}
\affiliation{Frankfurt Institute for Advanced Studies, Ruth-Moufang-Str. 1,
 60438 Frankfurt am Main, Germany}
\author{Andrey V. Solov'yov$\mathrm{^{b}}$}
\affiliation{Frankfurt Institute for Advanced Studies, Goethe University, 
Ruth-Moufang-Str. 1, 60438 Frankfurt am Main, Germany}
\affiliation{Virtual Institute on Nano Films (VINF), 
All\'{e}e des Noisetiers, 2 bte 30 Angleur, Belgium}


\begin{abstract}
A newly developed code, 
implemented as a part of
the \MBNExplorer package \cite{MBN_ExplorerPaper,MBN_ExplorerSite}
to simulate trajectories of an ultra-relativistic projectile in
a crystalline medium, is presented.
The motion of a projectile is treated classically by
integrating the relativistic equations of motion with account for
the interaction between the projectile and crystal atoms.
The probabilistic element is introduced by a random choice of
transverse coordinates and velocities of the projectile at the crystal
entrance as well as by accounting for the random positions of the atoms
due to thermal vibrations.
The simulated trajectories are used for
numerical analysis of the emitted radiation.
Initial approbation and verification of the code have been carried out
by simulating the trajectories and calculating the radiation emitted
by $\E=6.7$ GeV and $\E=855$ MeV electrons and positrons in oriented
Si(110) crystal and in amorphous silicon.
The calculated spectra are compared with the experimental data
and with predictions of the Bethe-Heitler theory
for the amorphous environment.
\end{abstract}

\maketitle

\section{Introduction} 
\label{Introduction}

In this paper we describe a new 
code for numerical simulation
of trajectories of ultra-relativistic electrons and positrons in
crystalline media.
The code is implemented as a part (a module) of the \MBNExplorer package
\cite{MBN_ExplorerPaper,MBN_ExplorerSite}.
The channeling motion of 855 MeV and 6.7 GeV electrons and positrons in
straight Si(110) crystal has been modeled by means of the code
and accompanied by calculations of the spectrum of emitted radiation.

The basic effect of the channeling process in a straight single crystal
is in an anomalously large distance which a particle can penetrate moving
along a crystallographic plane (planar channeling) or an axis
(axial channeling)
and experiencing the collective action of the electrostatic field of the
lattice ions \cite{Lindhard}.
The field is repulsive for positively charged particles and, therefore,
they are steered into the interatomic region, while negatively charged
projectiles move in the close vicinity of ion strings or planes.
Having introduced the continuum potential approximation for
the interaction of energetic projectiles and lattice atoms,
Lindhard  \cite{Lindhard}  demonstrated that a charged projectile can
move through the crystal following a particular crystallographic
direction if the incident angle is less than some critical value.

Under certain conditions
\cite{Tsyganov1976a,Tsyganov1976b} 
the guidance of channeled particles persists even if a crystal is bent.
In this case, the particle deviates from its initial direction
of motion due to extremely strong electrostatic field in the crystal.
The field strength is typically of the order of $10^{10}$ V/cm which
is equivalent to the magnetic field of approximately 3000 T.
Therefore, bent crystal can steer particles much more effectively than
any existing dipole magnet.

We refer to papers \cite{Lindhard,Gemmel,Sorensen1996,Baier,%
BiryukovChesnokovKotovBook,Uggerhoj_RPM2005,Uggerhoj2011}
which contain comprehensive reviews of theoretical and experimental
achievements in the investigation of the channeling effect in straight and
bent crystals as well of various related phenomena and
applications \footnote{Let us note that channeling can be discussed
not only for crystals but for any structured material which provides
``passages'', moving along which a projectile has much lower value of the
mean square multiple scattering angle than when moving along a random direction.
The examples of such materials are nanotubes and fullerites for which the
channeling effects has been also investigated \protect\cite{ArtruEtAl_PhysRep2005}.}.

Recently, the concept of a crystalline undulator (CU) was formulated
for producing undulator-like electromagnetic radiation in the hundreds of
keV up to the MeV photon energy range \cite{KSG1998,KSG_review_1999}.
In a CU, a beam of ultra-relativistic charged particles undergoes channeling
in a periodically bent crystal.
As a result, in addition to a well-known channeling
radiation \cite{ChRad:Kumakhov1976}, there appears the radiation
due to the undulating motion of channeling particles which follow the periodic
bending of crystallographic planes.
The intensity and characteristic frequencies of the CU radiation can
be varied by changing the type of channeling particles,
the beam energy, the crystal type  and the parameters of periodic bending.
Initially, it was proposed to use positron beams in CU \cite{KSG1998,KSG_review_1999}.
More recently, the feasibility of an electron-based CU has been demonstrated
\cite{PRL2007}.
The underlying fundamental physical ideas as well as the theoretical,
experimental and technological advances made during the last one and a half
decade in exploring various features of CUs and the emitted radiation
can be found in a recently published book
\cite{ChannelingBook2013}.

Several experimental attempts were made
\cite{BaranovEtAl_CU_2005,BaranovEtAl_CU_2006}
or planned to be made \cite{Backe_EtAl_2011a}
to detect the radiation from a positron-based CU.
So far, the attempts have not been successful due to
various reasons \cite{Backe_EtAl_2011a,ChannelingBook2013}.
However, quite recently the first signatures showing evidence for the
CU radiation were experimentally observed for 195--855 MeV electrons
at the Mainz Microtron (Germany) facility
\cite{Backe_EtAl_2010,Backe_EtAl_2011}.
The CUs, used in the experiment,
were manufactured in Aarhus University (Denmark) using the molecular beam
epitaxy technology to produce strained-layer Si$_{1-x}$Ge$_{x}$
superlattices with varying germanium content
\cite{MikkelsenUggerhoj2000,Krause_Diplom,Darmstadt01}.
Another set of experiments with diamond CUs is scheduled for the year
2013 at the SLAC facility (USA) with 10\dots 20 GeV electron
beam \cite{Uggerhoj_2012}.

Theoretical support of ongoing and future experiments as well as accumulation
of numerical data on channeling and radiative processes of ultra-relativistic
projectiles in crystals of various content and structure must be based on an
accurate procedure which allows one to simulate the trajectories
corresponding to the channeling and non-channeling regimes.
The procedure must include a rigorous description of the particle motion
and an efficient algorithm of its numerical realization.
It is strongly desirable to make the procedure as much universal and
model-independent as possible.
The universality implies applicability of the same code to simulate trajectories
of various projectiles (positively and negatively charged, light and heavy)
in an arbitrary scattering medium, either structured (straight, bent and
periodically crystals, superlattices, nanotubes etc) or amorphous (solids,
liquids).
The term ``model-independent'' implies that the only allowed parameters are
those which describe {\em pairwise} interactions (force-fields) of the
projectile with constituent atoms.

The existing codes, capable to simulate channeling process, do not comply
in full with the aforementioned conditions.
Some of them
\cite{Artru1990,Biryukov1995,Maisheev1996,Dechan01,Bogdanov_EtAl2010}
are based on the concept of the continuous potential \cite{Lindhard}.
This approximation, being adequate in describing the channeling motion, becomes
less accurate and more model-dependent when accounting for uncorrelated
scattering events.
The accurate description of the latter is essential for a quantitative analysis
of the dechanneling and rechanneling processes.
Other group of the channeling codes
\cite{BakEtal1985,SmuldersBoerma1987,Fomin_EtAl1997,ShulgaSyshchenko2005}
utilizes the scheme of binary collisions which
assumes that the motion of a projectile at all times is influenced by the
force due to the nearest atom.
Computer facilities available at present allow one to go beyond this limitation
and to account for the interaction with larger number of the crystal atoms.
Such an extension of the binary collisions algorithm was implemented in the
recent
code for electron channeling \cite{KKSG_simulation_straight,KKSG_NuovoCimento}.
The code, however, was based on the specific model for electron--atom scattering
which results in a noticeable overestimation of the mean scattering angle.
In more detail, this topic is addressed below in the paper.

To simulate propagation of particles through media, the channeling
process in particular, one can utilize approaches and algorithms used in modern
molecular dynamics (MD) codes
(a comparative review of codes can be found elsewhere \cite{MBN_ExplorerPaper}).
The latter allow one to model the dynamics of various molecular system by
efficient numerical integration of classical equations of motion for all
atoms in the system.
The interaction between atoms is implemented in terms of interatomic
potentials, the types and parameters of which can be chosen from a broad
range to ensure the most adequate quantitative description of the simulated
molecular system.
From this viewpoint, the MD concept can be applied to describe the motion of a
single projectile in the static field of atoms which constitute a scattering
medium.

However, to the best of our knowledge, no MD-based computer codes exist at
present that would permit simulation of the channeling phenomenon with atomistic
resolution.
This is mainly due to the following two reasons.
First, the charged projectile particles travel through the crystal in
an ultra-relativistic regime, and, therefore, their translocation
should be modeled with relativistic equations of motion,
which are typically not implemented in standard MD codes.
Second, the channeling phenomenon involves mesoscopically large crystals,
being $\mu$m-mm-cm in length, which cannot be handled using all-atom
MD approach.
To study the channeling phenomenon, we, therefore, have built a new MD-based
code that goes beyond the aforementioned drawbacks.
For this purpose we have used a recently developed \MBNExplorer software
package \cite{MBN_ExplorerPaper,MBN_ExplorerSite} and endowed
it with additional functionality.
\MBNExplorer was originally developed as a universal computer program to
allow investigation of structure and dynamics of molecular systems of
different origin on spatial scales ranging from nanometers and beyond
\cite{ISolovyov04,OObolensky05,ISolovyov03,ISolovyov08,JGeng08,Geng2010,%
Dick09,Dick10,Dick11}.
The general and universal design of \MBNExplorer code allowed us to expand
it's basic functionality with introducing a module that treats
classical relativistic equations of motion and generates the crystalline
environment dynamically in the course of particle propagation.
This module, combined with the variety of interatomic potentials implemented
in \MBNExplorer, makes the program a unique tool for studying relativistic
phenomena in various environments, such as crystals, amorphous bodies,
biological medium.
Below in the paper we introduce the key concepts and modifications done
in \MBNExplorer.

The channeling module, implemented currently in \MBNExplorer, aims at
efficient and reliable simulations of channeling of
ultra-relativistic projectiles in crystalline media.
Verification of the code against available experimental data
as well as against predictions of other theoretical models is a compulsory part
of our studies.
We have selected benchmark experimental
values $6.7$ GeV and $855$ MeV for the energy of projectile
electrons and positrons to simulate the trajectories and to calculate
spectral distribution of the emitted radiation
for two distinct environments:
Si (110) crystalline medium and amorphous Si.
The results of calculations for the $6.7$ GeV particles
are compared with the experimentally measured spectra
\cite{BakEtal1985,Uggerhoj1993}.
For amorphous silicon the numerical results are validated against
predictions of the Bethe-Heitler theory.

The paper is organized as follows.
In section \ref{Algorithm} we present the description of algorithms used 
to simulate the channeling process with \MBNExplorer
(\ref{MC_Simulations}) and to calculate the emission spectrum (\ref{wkb}).
Numerical results obtained for $6.7$ GeV and $855$ MeV electron/positron
channeling and emission spectra are discussed in section \ref{results}.
Concluding remarks are summarized in
section \ref{Conclusion}.
The paper has two appendices. 
In Appendix~\ref{Snapshot} we 
evaluate the accuracy of the model proposed
earlier \cite{KKSG_simulation_straight,KKSG_NuovoCimento}
for an ultra-relativistic electron--atom collision.
A collection of formulae related to the description of the bremsstrahlung
process within the framework of the Bethe-Heitler approximation is
presented in Appendix~\ref{BH}. 

\section{Description of the algorithm
\label{Algorithm}}

To perform 3D simulation of the propagation of ultra-relativistic
projectiles through a crystalline medium by means of \MBNExplorer the
following two additional features were to be added to the molecular
dynamic algorithms used in the package \cite{MBN_ExplorerPaper}.
The first feature concerns the implementation and integration of the
relativistic equations of motion.
The second one is the dynamic generation of the crystalline medium.
In more detail, these features are described in
section \ref{MC_Simulations} below.

The calculated dependencies of the coordinates $\bfr =\bfr(t)$ and velocities
$\bfv=\bfv(t)$ of the projectile on time
are used as the input data to generate the spectral and/or the
spectral-angular distributions of the emitted radiation.
These calculations are performed by means of the Fortran code (which is not
a part of \MBNExplorer) built upon the revisited algorithm
described earlier \cite{Krause_Diplom,Dechan01}.
The basic formalism is summarized below in
section \ref{wkb}.

\subsection{Simulations of the Channeling Process within \MBNExplorer}
\label{MC_Simulations}

Within the framework of relativistic classical mechanics\footnote{Discussion
on the range of validity of the classical approach to the channeling motion
of ultra-relativistic positive and negative projectiles one finds,
for example, in Ref. \cite{Uggerhoj_RPM2005}.}
the motion of an ultra-relativistic projectile of the charge $q$ and
mass $m$ in an external electrostatic field $\bfE(\bfr)$
is subject to the relativistic equations of motion which can be written
in the canonical form
\begin{eqnarray}
\begin{cases}
\dot{\bfp} = q\bfE \\
\dot{\bfr} = \bfv
\end{cases}\,.
\label{MC_Simulations.01}
\end{eqnarray}
A dot above a letter denotes differentiation with respect to time.
The momentum $\bfp$ written in terms of velocity reads $\bfp = m\gamma \bfv$
where $\gamma$ stands for the Lorentz factor $\gamma=(1-v^2/c^2)^{-1/2}=\E/mc^2$
with $\E$ being the projectile energy.

The differential equations (\ref{MC_Simulations.01}) are
to be integrated for $t\geq 0$ using the
initial values of the coordinates $(x_0,y_0,z_0)$ and the velocity components
$(v_{x0},v_{y0}, v_{z0})$ of the particle.
To ensure an accurate numerical integration the fourth-order Runge-Kutta
scheme has been implemented.

In application to the particle motion in a scattering medium (a crystal,
in particular), the important issue is an accurate and efficient computation of
the electrostatic field due to the medium atoms.
In the current version of the channeling module of \MBNExplorer
the electrostatic potential $U(\bfr)$ is represented as a sum of atomic
potentials $\Uat$
\begin{eqnarray}
U(\bfr)
=
\sum_{j} \Uat\left(\rho_j\right)\Bigl|_{\brho_j=\bfr - \bfR_j },
\label{MC_Simulations.02}
\end{eqnarray}
where $\bfR_j$ stands for the position vector of the $j$-th atom.
The force acting on the projectile at point $\bfr$ is calculated as
$\bfF = q\bfE(\bfr) = -q\bnabla_{\bfr}U(\bfr)$ .

A number of approximate methods have been developed to construct simple
analytical representations of atomic potentials
(see, for example, Ch. 9.1 in Ref. \cite{Baier}).
All these schemes can be straightforwardly added to the library of pairwise
potentials used in \MBNExplorer.
The current version of the package utilized the widely used
Moli\`{e}re approximation \cite{Moliere} as well as more recent
approximation suggested by Pacios \cite{Pacios1993}.
For further referencing let us reproduce the Moli\`{e}re formula
for the electrostatic potential of a neutral atom:
\begin{eqnarray}
U_{\rm M}(\rho)
=
{Ze \over \rho}\, \chi(\rho),
\qquad
\chi(\rho)
=
\sum_{j=1}^3 \alpha_j \, \ee^{-\beta_j \rho /\aTF}\,,
\label{MC_Simulations.03}
\end{eqnarray}
where $Z$ is the atomic number.
The Thomas-Fermi radius $\aTF$ is related to the Bohr radius $a_0$ via
$\aTF = {0.8853 Z^{-1/3}}\, a_0$.
The coefficients in the screening function $\chi(\rho)$ read as:
$\alpha_{1,2,3} = (0.35; 0.55; 0.1)$ (so that $\sum_{j=1}^3 \alpha_j=1$)
and  $\beta_{1,2,3} = (0.3; 1.2; 6.0)$.

In recent studies \cite{KKSG_simulation_straight,KKSG_NuovoCimento}
another model for an ultra-relativistic projectile--atom
interaction was suggested.
The underlying idea of the model is that due to high speed of the projectile the
interaction interval is short enough to substitute the atom with its ``snapshot''
in which the atomic electrons are seen as point-like charges distributed
around the nucleus.
In Appendix~\ref{Snapshot} we demonstrate that such a 
``snapshot'' approximation overestimates the mean scattering angle in a single 
projectile--atom collision.

Formally, the sum in Eq.~(\ref{MC_Simulations.02}) is carried out over all atoms
of the sample.
Taking into account that $\Uat\left(\rho_j\right)$ decreases
rapidly at the distances $\rho_j\gg \aTF$ from the nucleus,
one can introduce the cutoff $\rho_{\max}$ above which the
contribution of $\Uat\left(\rho_j\right)$ is negligible.
Therefore, for given observation point $\bfr$ the sum can be restricted
to those atoms which are located inside the sphere of the radius $\rho_{\max}$.
To facilitate the search for such atoms the linked cell algorithm,
implemented in \MBNExplorer, is employed.
The algorithm implies (i) a subdivision of the sample into cubic cells of a
 smaller size, and (ii) an assignment of each atom to a certain cell.
Choosing the cell size equal to $\rho_{\max}$ one restricts the sum to
those atoms from the cell containing the observation point and from the
26 neighbouring cells which lie inside the cutoff sphere.
As a result, the total number of computational operations can be reduced
considerably.

The described scheme is used to calculate the force $q\bfE$ acting on the projectile
at each integration step in (\ref{MC_Simulations.01}).
In section \ref{results} we present the results of simulation of the channeling
process of ultra-relativistic electrons and positrons in crystalline silicon.
In this case we used $\rho_{\max}= 5$ \AA\, which enters Eq.~(\ref{MC_Simulations.03}), 
and for a Si atom is to be compared to the value $\aTF=0.194$ \AA.
The lattice constant of a cubic Bravais cell of a silicon crystal is 5.43 \AA.
Each unit cell contains eight atoms.
Therefore, at each integration step the sum on Eq.~(\ref{MC_Simulations.02})
was carried out on average for over 30 atoms.

To simulate the channeling motion along a particular crystallographic
plane  with Miller indices $(klm)$ the following algorithm has been used.

\begin{figure}[!t]
\centering
\includegraphics[width=13cm,clip]{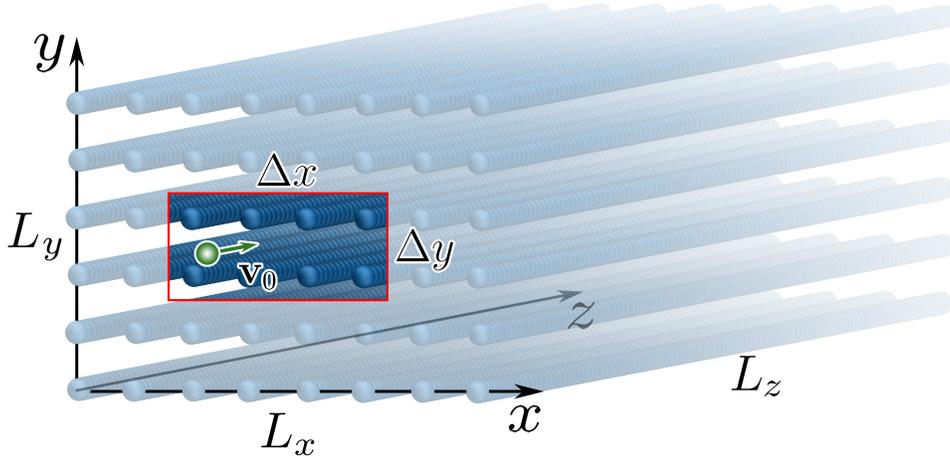}
\caption{The crystalline structure is generated inside the simulation box
with the dimensions $L_{x,y,z}$ along the chosen coordinate axes.
The $z$-axis is aligned with the incident beam direction and is parallel to the
crystallographic plane along which the channeling is to be simulated.
The $y$-axis is perpendicular to the plane.
The lattice nodes are generated in accordance with
Eq. (\ref{MC_Simulations.04}).
At the entrance the $x$ and $y$ coordinates of the particle are randomly chosen
to lie in the central part of the $(xy)$ plane (the highlighted rectangle
with the sides $\Delta x,\Delta y$).
The initial velocity $\bfv_0$ is predominantly oriented along $z$.
See also explanations in the text.}
\label{MC_Simulations:fig.01}
\end{figure}

As a first step, a crystalline lattice is generated inside the simulation box
(parallelepiped) of the size $L_x\times L_y \times L_z$
illustrated in Fig.~\ref{MC_Simulations:fig.01}.
The $z$-axis is oriented along the beam direction and is parallel to
the $(klm)$ plane.
To avoid the axial channeling (when not desired) the $z$-axis
must not be collinear with major crystallographic axes.
The $y$-axis is perpendicular to the plane.
The position vectors of the nodes are generated according to the rule:
\begin{eqnarray}
\bfR_i^{(0)}(k_x,k_y,k_z)
=
\bfT(k_x,k_y,k_z)
+
\bfP_i
&\equiv&
k_x\bfa_x + k_y \bfa_y + k_z\bfe_z
+
\bfP_i,
\quad i=1,2,\dots, n\,.
\label{MC_Simulations.04}
\end{eqnarray}
Here, $\bfa_{x,y,z}$ are the lattice vectors and
$k_{x,y,z}$ are integers.
Thus, the transition vector $\bfT(k_x,k_y,k_z)$
defines the position of a unit cell.
The vector
$\bfP_i = \varkappa_{ix} \bfa_x + \varkappa_{iy}\bfa_y + \varkappa_{iz}\bfa_z$
with $\varkappa_{ix,y,z} \in [0,1]$ defines the position of the $i$-th node
(out of the total number $n$) in the unit cell.

To illustrate the latter step, let us consider a diamond-type lattice which
describes  diamond, silicon and germanium crystals.
In this case the three vectors $\bfa_{x,y,z}$, being orthogonal, are of the
same length $a$ which defines the lattice constant.
Its values (at $T=300$ K) for diamond, Si and Ge are
3.567, 5.431 and 5.646 \AA, respectively.
Each unit cell contains 8 atoms, the position vectors
of which are
\begin{eqnarray}
\!\!\!\!\!\!
\begin{array}{lllll}
&\bfP_1 = (0,0,0) a, &
\bfP_2 = \left(0,{1\over 2},{1\over 2}\right) a, &
\bfP_3 = \left({1\over 2},0, {1\over 2}\right)a, &
\bfP_4 = \left({1\over 2},{1\over 2},0\right) a,
\\
&\bfP_5 = \left({1\over 4},{1\over 4},{1\over 4}\right) a, &
\bfP_6 = \left({1\over 4},{3\over 4},{3\over 4}\right) a, &
\bfP_7 = \left({3\over 4},{1\over 4},{3\over 4}\right) a, &
\bfP_8 = \left({3\over 4},{3\over 4},{1\over 4}\right) a\,.
\end{array}\!\!\!
\label{Diamond_Si_Ge.02}
 \end{eqnarray}

Once the position vectors $\bfR_j^{(0)}$ ($j=1,2,\dots, N$) for all nodes inside
the simulation box are defined,
the position vectors $\bfR_j = \bfR_j^{(0)} + \bDelta_j$ of the
atomic nuclei are generated.
This is done with account for the thermal vibrations which result in random
displacement $\bDelta_j$ from the nodal positions.
Each component of $\bDelta_j$ is normally distributed
\begin{eqnarray}
w(\Delta_{jk})
=
{1\over \sqrt{2\pi u_T^2}} \exp\left(- {\Delta_{jk}^2 \over 2u_T^2}\right),
\quad
k=x,y,z.
\label{MC_Simulations.05}
\end{eqnarray}
Here $u_T$ is the root-mean-square amplitude of thermal vibrations.
The numerical results for a Si crystal presented below in
section \ref{results}
were obtained for $u_T=0.075$ \AA\, which corresponds to the room temperature
\cite{Gemmel}.

Integration of the equations of motion, Eqs.~(\ref{MC_Simulations.01}),
starts at $t=0$ when the particle ``enters'' the crystal at $z=0$.
The initial coordinates $x_0$ and $y_0$ are randomly chosen
to be lying in the central part of the $(xy)$-plane of the sizes $\Delta x = 2d$,
$\Delta y = d$ where $d$ is the interplanar distance for the $(klm)$ planes, see
Fig. \ref{MC_Simulations:fig.01}.
The initial velocity $\bfv_0=(v_{0x},v_{0y},v_{0z})$ is predominantly oriented
along $z$, i.e. the conditions $v_{0z}\approx c \gg v_{0x},v_{0y}$ are implied.
The transverse components $v_{0x}, v_{0y}$ can be chosen with account for the
beam emittance.

\begin{figure}[!t]
\centering
\includegraphics[width=13cm,clip]{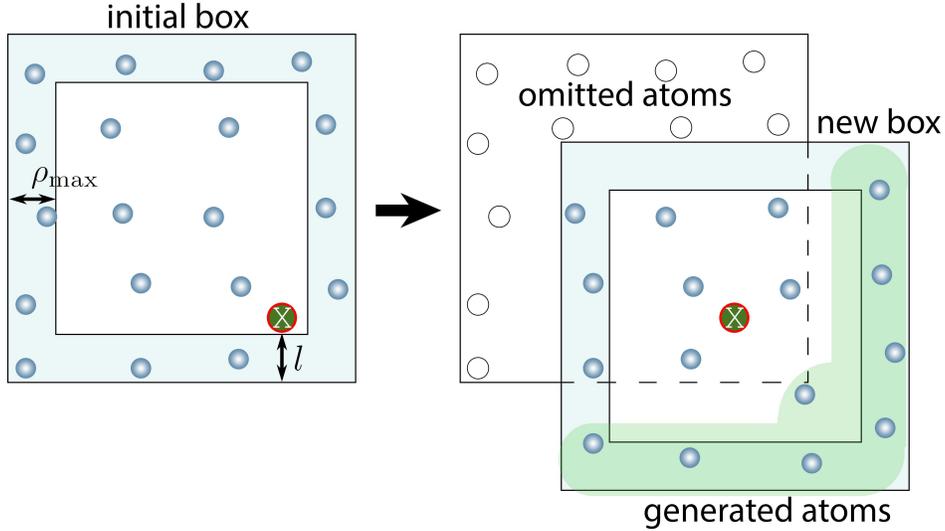}
\caption{Illustration of the dynamic simulation box algorithm.
When an X-marked projectile approaches the face of the
initial simulation box (left panel) by a distance $l\approx \rho_{\max}$ a
new simulation box of the same size is generated (right panel)
with the particle placed approximately in its geometrical center.
The \textcolor{red}{position of the} atoms (small shadowed circles)
located in the intersection of the old and
the new boxes are not changed.
In the rest part of the new box the atomic positions
are generated anew as described in the text.}
\label{MC_Simulations:fig.02}
\end{figure}

To simulate the propagation of a particle through a crystal of finite thickness
$L$ a new type of boundary conditions, the so-called
``dynamic simulation box'', has been implemented in \MBNExplorer.
This algorithm, illustrated in Fig. \ref{MC_Simulations:fig.02},
implies the following.

The projectile moves within the simulation box interacting with the atoms lying
inside the cutoff sphere.
To optimize the numerical procedure the lengths $L_{x,y,z}$ are chosen to
be larger than $\rho_{\max}$ by a factor of $3\dots 5$.
Once the distance $l$ from the projectile to the nearest face becomes ``nearly''
equal to $\rho_{\max}$ a new simulation box of the same size is generated
with its geometrical center coinciding (approximately) with the
position of the projectile.
To avoid spurious change in the force $q\bfE$ acting on the projectile
the positions of the atoms located in the intersection of the old and the new
simulation boxes are not changed.
In the rest part of the new box the positions of
atomic nuclei are generated following the scheme described
through Eqs.~(\ref{MC_Simulations.04}) and (\ref{MC_Simulations.05}).
The simulation is interrupted when the $z$ coordinate of the particle
becomes equal to the crystal thickness $L$.

Using the described algorithm we have simulated a number of trajectories
for electrons and positrons of the energies
$\E=855$ MeV and $6.7$ GeV moving along the (110) planes in straight
silicon crystals.
The obtained results are presented and discussed in more detail in
section \ref{results}.
The motion in the amorphous silicon has been also simulated.
For doing this it is necessary to avoid incidental alignment of the initial
velocity $\bfv_0$ with major crystallographic directions.
We used this regime to calculate the spectral and spectral-angular
distribution of the incoherent bremsstrahlung.

\subsection{Quasi-Classical Formalism for the Radiated Energy
\label{wkb}}

In many cases, the motion of an ultra-relativistic particle, moving in an external
field, can be treated within the framework of classical mechanics.
The applicability of the classical description is subject to the
condition that the relative variation of the de Broglie wavelength
$\lambda_{\rm B}= h/p$ of the projectile must be small over the distances
of the order of $\lambda_{\rm B}$.
This condition, written in terms of the maximum gradient $\dUmax$
of the external field, reads:
$m\hbar \dUmax/p^3 \ll 1$, where $m$ and $p\approx \E/c$ are projectile's mass
and momentum.
Taking into account that $\dUmax\sim 10^{1}\dots10^{2}$ GeV/cm for a planar
crystalline potential and by approximately an order of magnitude higher for
an axial potential\cite{Baier},
one demonstrates that the condition is
well-fulfilled for projectile positrons and electrons with
$\E \sim 10^2$ MeV and higher.

The process of photon emission can be treated classically provided the
photon energy  is small compared to $\E$:  $\hbar\om/\E \to 0$.

If both of the aforementioned conditions are met, one can
calculate the spectral-angular distribution the radiated energy using the
standard formulae of classical electrodynamics
\cite{Landau2,Jackson}.

The main drawback of the classical framework is that it does not allow
a self-consistent description of the radiative recoil,
i.e. the change of the projectile energy due to the photon emission.
As a result, purely classical description fails when the
ratio $\hbar \om /\E$ is not infinitesimally small.

An adequate approach to the radiation emission by ultra-relativistic
projectiles in the (nearly) whole range of the photon energies
was developed by Baier and Katkov in the
late 1960th \cite{Baier67}
and was called by the authors the ``operator quasi-classical method''.
The details of the formalism, as well as  its application to a variety
of radiative processes, can be found elsewhere~\cite{Baier,Baier1,Landau4}.

A remarkable feature of this method is that it allows one to combine
the classical description of the motion in an external field
and the quantum effect of radiative recoil.

The classical description of the motion is valid provided
the characteristic energy of the projectile in an external field,
$\tilde{\E}_0$,
is much less than its total energy,
$\E = m \gamma c^2$.
The relation
$\tilde{\E}_0/\E \propto \gamma^{-1} \ll 1$
is fully applicable in the case of an ultra-relativistic projectile.
The quasi-classical approach neglects the terms
$\tilde{\E}_0/\E$
but explicitly takes into account the quantum corrections due to the radiative recoil.
The method is applicable in the whole range of the
emitted photon energies, except for the extreme high-energy tail of the spectrum
$\left(\E-\hbar\om\right) /\E \ll 1$.

Within the framework of Baier and Katkov quasi-classical formalism the
energy radiated within the cone $\d \Om = \sin\theta\d\theta\,\d\phi$
by an ultra-relativistic particle moving along the trajectory $\bfr=\bfr(t)$
is written as
\begin{eqnarray}
{ \d^2 E \over \d(\hbar\om)\, \d \Om}
=
\alpha \,
{ q^2\omega^2  \over 8 \pi^2 }
\int\limits_{-\infty}^{\infty} \d t_1
\int\limits_{-\infty}^{\infty} \d t_2\,
\ee^{\i \,\omega^{\prime} \left(\psi(t_1) -\psi(t_2)\right)}
\left[
\left( 1+(1+u)^2 \right)
\left(
\bfbeta_1\cdot\bfbeta_2  -1
\right)
+
{u^2 \over \gamma^2}
\right],
\label{wkb:eq.01}
\end{eqnarray}
where $\alpha= e^2/ \hbar\, c$ is the fine structure constant,
$q$ is the charge of a projectile in units of the elementary
charge, $\bfbeta_{1,2} =\bfv(t_{1,2})/c$ denote the velocities, scaled by $c$,
at time instants $t_1$ and $t_2$.
The phase function reads
$\psi(t) = t - \bfn\cdot\bfr(t)/ c$
where $\bfn$ is the unit vector in the direction of the photon emission.

The quantities $\omega^{\prime}$ and $u$ account for the radiative recoil:
\begin{eqnarray}
\omega^{\prime}
=
(1+u)\, \om,
\qquad
u =
{\hbar \om \over \E - \hbar \om}.
\label{wkb:eq.03}
\end{eqnarray}
In the classical limit
$u\approx \hbar\om /\E\to 0$ and $\om^{\prime} \to \om$, so that
Eq. (\ref{wkb:eq.01}) reduces to the classical formula \cite{Landau2,Jackson}.

\begin{figure}[!t]
\centering
\includegraphics[width=13cm,clip]{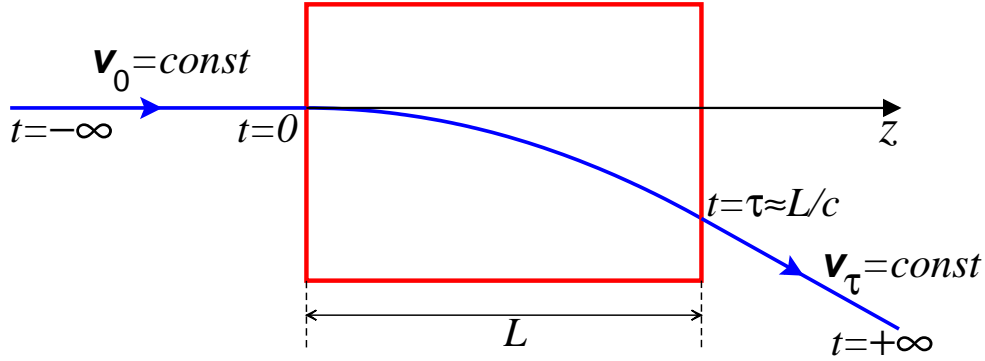}
\caption{A trajectory of the ultra-relativistic particle ($v\approx c$)
which experiences the action of the external field within the scattering medium of the thickness
$L$. Before entering the medium, i.e. within the time interval $t<0$, and after
leaving it at $t=\tau\approx L/c$
the particle moves with constant velocities along the straight lines.
Inside the medium the motion is subject to the forces acting on the particle.}
\label{wkb:fig.01}
\end{figure}

The trajectory of a particle which propagates through a crystalline or amorphous
medium of the thickness $L$ can be divided into three segments, as illustrated
in Fig. \ref{wkb:fig.01}.
Below the following motion of the projectile will be considered.
\begin{itemize}
\item
Within the time intervals $t=[-\infty,0]$ and $t=[\tau,\infty]$ the
projectile moves outside the medium along the straight lines
with constant velocities $\bfv_0$ and $\bfv_{\tau}$, respectively.
The quantity $\tau$ is the time of flight through a spatial domain of
thickness $L$.
In the case of ultra-relativistic projectile and assuming
small scattering angle limit one estimates the time of flight as
$\tau \approx L/c$.

\item
During the interval $t=[0,\tau]$ the particle experiences the action of the
external field and, as a result, moves along some non-linear trajectory defined by
$\bfr=\bfr(t)$.

\end{itemize}

Assuming the relativistic factor to satisfy
a strong inequality $\gamma \gg 1$,
one expands the phase function $\psi(t)$ and the rest of integrand
in (\ref{wkb:eq.01}) in powers of $\gamma^{-1}$
retaining the dominant non-vanishing terms.
After some algebra\cite{Baier,ChannelingBook2013}
one represents the spectral-angular distribution
in the following form, which is convenient
for numerical evaluation:
\begin{eqnarray}
\nonumber
{ \d^2 E \over \hbar\d\om\, \d \Om}
&=&
\alpha q^2\omega^2\,
{   (1+u) (1 + \Delta)\over 4 \pi^2 }
\left[
{\Delta \, \left|S_z\right|^2\over \gamma^2 (1 + \Delta)}
+
\left| \sin\phi  S_{x} - \cos\phi S_{y}\right|^2
\right.\\
\label{wkb:eq.04}
&&
+\left|
\theta S_z
-
 \cos\phi  S_{x}
- \sin\phi  S_{y}
\right|^2
\Biggr],
\end{eqnarray}
where $\Delta = {u^2 /2(1+u)}$.
The quantities $S_{x,y,z}$ are defined as
\begin{eqnarray}
S_{z}
&=&
\int\limits_{-\infty}^{\infty} \d t\,
\ee^{\i\, \omp \psi(t)}
=
\int\limits_{0}^{\tau}
\d t\, \ee^{\i\,\omp \psi(t)}
-
{\i\over \omp}
\left(
{\ee^{\i\, \omp  \psi(0) }
\over  D_{0}}
-
{\ee^{\i\,\omp \psi(\tau)}
\over D_{\tau}  }
\right),
\label{wkb:eq.05}
\\
S_{x,y}
&=&
\int\limits_{-\infty}^{\infty} \d t\,
\beta_{x,y} (t)
\ee^{\i\omp \psi(t)}
=
\int\limits_{0}^{\tau}
\d t\,
\beta_{x,y}(t)
 \ee^{\i\,\omp \psi(t)}
-
{\i\over \omp}
\left(
\beta_{0 x,y}
{\ee^{\i\omp  \psi(0) }
\over   D_{0}}
-
\beta_{\tau x,y}
{\ee^{\i\omp \psi(\tau)}\over  D_{\tau} }
\right),
\label{wkb:eq.06}
\end{eqnarray}
where $D_{0,\tau}=1 -  \bfn\cdot\bfbeta_{0,\tau}$.

The right-hand side of Eq. (\ref{wkb:eq.04})
is written in the limit of small emission
angles $\theta\ll 1$ with respect to the initial velocity $\bfv_0$
which defines the $z$-direction, see Fig. \ref{wkb:fig.01}.
In this limit the phase function $\psi(t)$ reads as
\begin{eqnarray}
\psi(t)
&=
t - {\bfn\cdot\bfr(t) \over c}
\approx
t -  \left(1-{\theta^2 \over 2}\right) {z\over c}
-\theta\, {x\cos\phi + y\sin\phi \over c}\,.
\label{wkb:eq.07}
\end{eqnarray}

The non-integral terms on the right-hand sides of  (\ref{wkb:eq.05})
and  (\ref{wkb:eq.06}) are due to the motion along the initial and final straight
segments of the trajectory.
{Thus, within the framework of quasi-classical approach,
Eqs. (\ref{wkb:eq.04})--(\ref{wkb:eq.06}) explicitly take into account
the dependence of the spectral-angular distribution on the thickness $L$
of a scattering medium which enters the formulae via the time-of-flight
$\tau\approx L/c$.}

Numerical evaluation of the integral terms by means of any classical formula
based on the sequence of time instants $t_1=0, t_2, t_3, \dots, t_N=\tau$ is stable
only when the following strong inequality is met:
\begin{eqnarray}
|\om\Delta\psi(t)|
=
\left|\om\Bigl(\psi(t+\Delta t) - \psi(t)\Bigr) \right|
\approx
\om\left| {\d\psi \over \d t}\right| \Delta t
\ll 1\,.
\label{TimeStep.02}
\end{eqnarray}
Here $\Delta t=t_{j+1}-t_i>0$ is the time step used for the integration over the
interval $[t_j,t_{j+1}]$.

In the limit of small emission and scattering  angles
the derivative ${\d\psi /\d t}$ can be transformed and estimated
as
\begin{eqnarray}
{\d\psi(t) \over \d t}
&=&
1 - \bfn\cdot\bfbeta
\approx
{1\over 2\gamma^2}
+
{1\over 2}
\left(
\theta^2 + \theta_{\bfv}^2
-
2\theta\, \theta_{\bfv}\, \cos(\phi-\phi_{\bfv})
\right)
\leq
{1\over 2}
\left[
{1\over \gamma^2}
+
\left(\theta + \theta_{\bfv}\right)^2
\right],
\label{TimeStep.08}
\end{eqnarray}
where $(\theta_{\bfv},\phi_{\bfv})$ are the scattering angles measured with respect
to the initial velocity.

Hence, the step of numerical integration must be chosen to satisfy the condition
\begin{eqnarray}
\Delta t
\ll
{2\gamma^2 \over \om}
\left[1 + \gamma^2
\left(
\theta + \theta_{\bfv}
\right)^2
\right]^{-1}\,.
\label{TimeStep.10}
\end{eqnarray}

\section{Numerical Results}
\label{results}

Channeling of charged particles in crystals is accompanied by the
channeling radiation \cite{ChRad:Kumakhov1976}.
This specific type of electromagnetic radiation arises due to the
transverse motion of the particle inside the channel under
the action of the planar or axial field~--~the channeling oscillations.

A considerable amount of experimental data has been accumulated during last
decades on the characteristics of channeling radiation emitted by
GeV and multi-GeV electrons and positrons in strong crystalline fields
\cite{Andersen_ChanRadReview_1983,BakEtal1985,BakEtal1988,Uggerhoj1993,Instrum,%
Medenwaldt1991,BaurichterEtAl1997,Berman_Nato2001_review,%
KirsebomEtAl_NIMB_2001,Uggerhoj_RPM2005,%
Channeling2008,Uggerhoj2011}.
More recent activity includes experiments with sub-GeV high-quality electron beam
carried out at the MAinz MIcrotron (MAMI)
\cite{Backe_EtAl_2008,Backe_EtAl_2010a,Backe_EtAl_2010,Backe_EtAl_2011,Backe_EtAl_2011a}.
One of the goals of these ongoing experiments is to test
theoretical prediction on the feasibility of an electron-based
crystalline undulator \cite{PRL2007,ChannelingBook2013}.

The verification of the developed
code against available experimental data as well as against predictions of
other theoretical models is an important part of our studies.
To this end, we have selected benchmark experimental values $6.7$ GeV and $855$ MeV 
for the energy of projectile electrons and positrons.
The projectile trajectories and the spectral distribution of emitted radiation
have been computed for (110) crystalline medium and for amorphous Si.
The results of calculations for the $6.7$ GeV particles
are presented in section \ref{6_7GeV_Si} and are compared with the
experimentally measured spectra \cite{BakEtal1985,Uggerhoj1993}.
For amorphous silicon the numerical results are validated against
predictions of the Bethe-Heitler theory
(see Appendix~\ref{BH}).
Section \ref{855MeV_Si} presents the results of calculations for
$\E=855$ MeV electrons and positrons.

\subsection{Results for 6.7 GeV electrons and positrons}
\label{6_7GeV_Si}

\begin{figure}[!t]
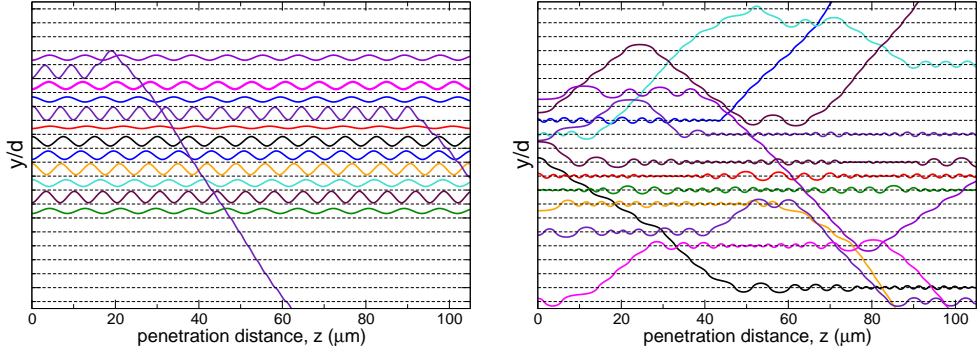

\centering
\includegraphics[scale=0.27,clip]{Figure_04left.eps}%
\hspace*{0.4cm}
\includegraphics[scale=0.27,clip]{Figure_04right.eps}%
\caption{Channeling of 6.7~GeV positrons (left) and electrons (right)
in a $105$ $\mu$m thick silicon crystal.
The plots show typical trajectories of the particles initially collimated
along Si(110) crystallographic planes.
Horizontal dashed lines indicate the planes separated by the
distance $d=1.92$ \AA.}
\label{trajs_6_7GeV_el_pos.fig}
\end{figure}

For the Si(110) planar orientation, both positrons and electrons exhibit
channeling motion as it is illustrated in Fig.~\ref{trajs_6_7GeV_el_pos.fig}
by sets of typical simulated trajectories.

For positrons, noticeable are nearly harmonic oscillations between the
neighbouring planes.
This is in accordance with a well-known result established within the
framework of the continuum model of channeling (see, e.g., Ref. \cite{Gemmel}).
Indeed, for a positively charged projectile the interplanar potential can be
approximated by parabola in most part of the Si(110) channel.
Therefore, the channeling oscillations are very close to the harmonic type.
Another feature of  positron channeling through a $L=105$ $\mu$m thick crystal
is a small number of the dechanneling events (the two examples presented in
the figure were found in forty randomly chosen trajectories).
This is also not surprising if one compares the crystal size with the dechanneling
length $\Ld\approx 0.4$ mm for a 6.7 GeV positron in Si(110).
The latter value can be obtained using Eq. (1.50) from the
book \cite{BiryukovChesnokovKotovBook}
with the correction for a light projectile
introduced later \cite{Dechan01}.

Much less regular are the channeling oscillations of electrons, see
Fig.~\ref{trajs_6_7GeV_el_pos.fig} (right).
In contrast to positrons, the electron trajectories exhibit a broader variety of
features: channeling motion, over-barrier motion,
rechanneling process, rare events of hard collisions etc.
First, let us note that the dechanneling length of a 6.7 GeV electron in Si(110),
estimated with the help of Eq. (10.1) from the book \cite{Baier},
is $\Ld\approx 130$ $\mu$m.
Therefore, it is not surprising that a noticeable fraction of electrons,
although channeling in close vicinity to the plane, stays in the channeling
mode from the entrance point up to the end of the crystal.
The events of rechanneling, i.e., capture to the channeling mode of an
over-barrier particle, are quite common for electrons.
Even the multiple rechanneling events are not rare.
This phenomenon has been already noted in recent 
simulations of the electron channeling \cite{KKSG_simulation_straight}
with a qualitative explanation provided of the difference in the
rechanneling rate for positively and negatively charged projectiles.
The conclusion drawn on the much lower rechanneling probability for a positron
than that for an electron is clearly illustrated by comparing the trajectories
on the left and right panels of the figure.
Also it is worth noting a visible anharmonicity in the channeling oscillations of
electrons which is a direct consequence of a strong deviation of
the electron interplanar potential from a harmonic shape \cite{BakEtal1985}.
As a result, the period of the oscillations varies with the amplitude.

\begin{figure}[!t]
\centering
\includegraphics[width=\textwidth,keepaspectratio=true]{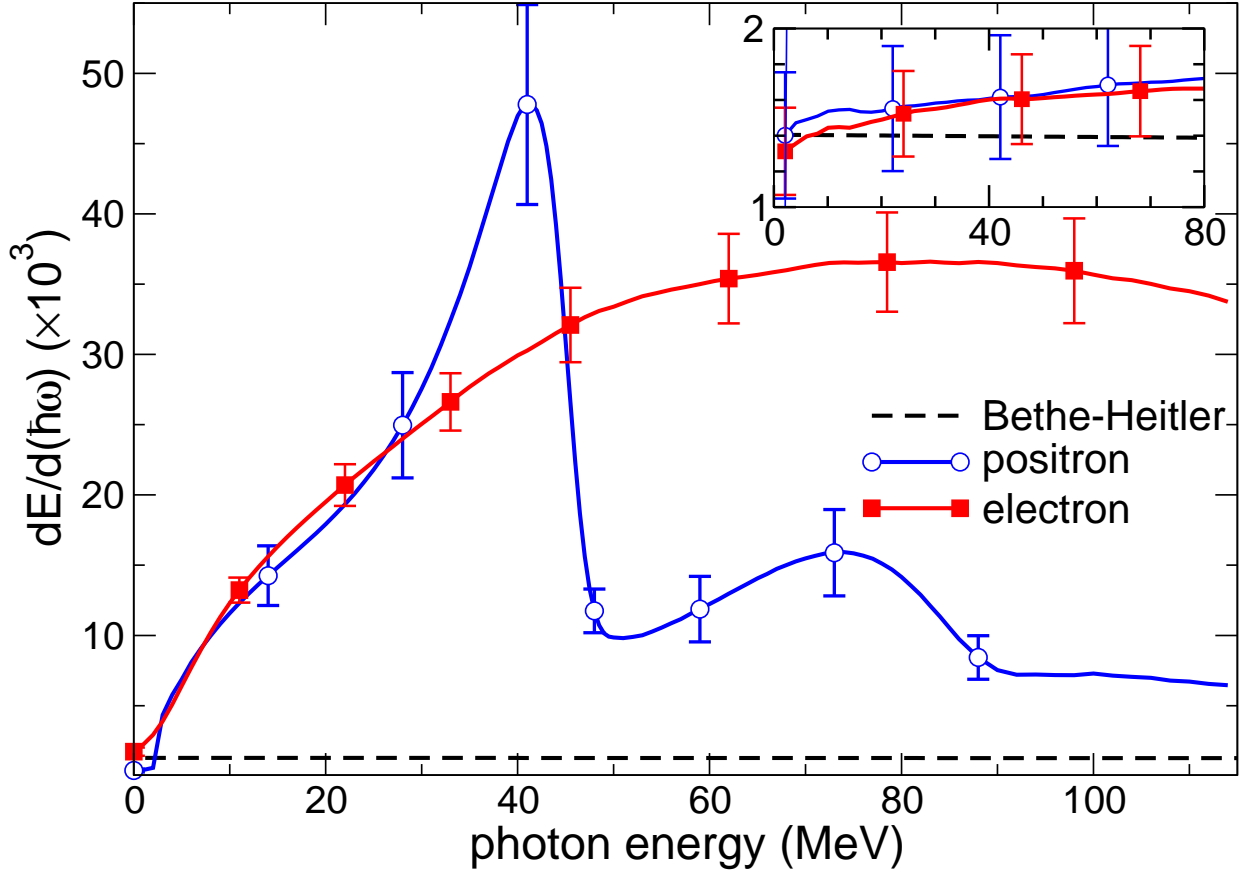}
\caption{Radiation spectra from
6.7~GeV positrons and electrons
(as indicated) channeling through a $105$ $\mu$m thick Si(110).
Dashed black line shows the Bethe-Heitler spectrum in amorphous silicon.
The inset presents the spectra calculated for the simulated trajectories
in amorphous Si.
{The solid curves are drawn over ca 200 photon energy points in which the spectra
were calculated.
The symbols (circles and rectangles) mark a small fraction of the points
and are drawn to illustrate typical statistical errors
(due to a finite number of the simulated
trajectories) in different parts of the spectrum.}
}
\label{dE_6_7GeV_el_pos_v04.fig}
\end{figure}

The simulated trajectories were used to calculate spectral distribution
of the emitted radiation.
The solid curves in Fig.~\ref{dE_6_7GeV_el_pos_v04.fig} represent the
dependencies  $\d E/\d(\hbar\om)$ calculated for 6.7 GeV electrons and positrons
aligned along Si(110) crystallographic plane at the crystal entrance.
Statistical uncertainties due to the finite number ($\approx 500$ in each case)
of the analyzed trajectories are indicated by the error bars
(the confidence interval)
which correspond to the probability $\alpha=0.999$.
The spectra were computed for a detector aperture of $\theta_{\rm a} =0.35$ mrad
hinted by the description of the experiments \cite{BakEtal1985,Uggerhoj1993}.
This value exceeds the ``natural'' emission cone $\gamma^{-1}$ by a factor
of $\approx5$.
Therefore, the calculated curves account for nearly all emitted radiation.

First, we note that for both electrons and positrons the intensity of radiation in
the oriented crystal greatly exceeds (by more than an order of magnitude)
that by the same projectile in an amorphous medium.
The latter is indicated by the dashed line and was calculated within the
framework of the Bethe-Heitler approach using Eqs. (\ref{BH:eq.08}) and
(\ref{BH:eq.09}).
The enhancement is due to the contribution to $\d E/\d(\hbar\om)$ coming from
the particles moving along quasi-periodic channeling trajectories,
which bear close resemblance with the undulating motion.
As a result, constructive interference of the waves emitted from different
but similar parts of the trajectory increases the intensity.
For each value of the emission angle $\theta$ the coherence
effect is most pronounced for the radiation into harmonics,
which frequencies can be estimated as follows \cite{Baier}:
\begin{eqnarray}
\om_{n}
=
{2\gamma^2\, \Om_{\rm ch}
\over
1 + \gamma^2 \theta^2 + K_{\rm ch}^2/2}
\, n ,
\quad
n=1,2,3,\dots \,,
\label{Section2.05_2:eq.01}
\end{eqnarray}
where $\Om_{\rm ch}$ is the frequency of channeling oscillations
and $K_{\rm ch}^2 = 2\gamma^2 \left\langle v_{\perp}^2\right\rangle/c^2$
is the mean square of the undulator parameter related to them.
Within the framework of continuous potential approximation, these quantities
are dependent on the magnitude of the transverse energy which, in turn,
determines the amplitude of oscillations.
The only exception is the  harmonic potential for which $\Om_{\rm ch}$ is
independent on the amplitude.

Different character of channeling by positrons and electrons results
in differences in the spectra of the channeling radiation.

The nearly perfect sine-like channeling trajectories of positrons
lead to the emission spectrum close to that of the undulator radiation with
$K^2 < 1$ \footnote{Using Eq. (B.5) from the book
\protect\cite{ChannelingBook2013},  one estimates
$K^2_{\rm ch}\approx 0.4$.}.
A pronounced peak in the photon energy range $20\dots 45$ MeV
is due to the emission in the fundamental harmonic ($n=1$).
The maximum corresponds to the forward emission ($\theta=0$) and can be
estimated from (\ref{Section2.05_2:eq.01}) as $\hbar\om\approx 40$ MeV.
The second, less accented peak, corresponds to the emission in
the second harmonic.

In contrast, in the electron spectrum the undulator effect is completely
smeared out due to strong anharmonicity of the channeling trajectories.

In addition to the channeling spectra we have computed the spectra for amorphous
Si medium.
For doing this, the trajectories of electrons and positrons were simulated
for a random orientation of the crystal with the care taken to avoid
major crystallographic directions along the beam axis.
The spectral-angular distributions of the simulated radiation were integrated
over $\theta_{\rm a} =0.4$ mrad aperture.
The calculated spectra are compared in the inset of
Fig.~\ref{dE_6_7GeV_el_pos_v04.fig}.
Remarkably, the spectra produced by positrons and electrons in amorphous Si
appeared to practically coincide with each other and to agree quite well with
the Bethe-Heitler result.
We consider this agreement as indicating the reliability of our numerical
simulations.
{As known (see, for example, Section 7 in Ref. \cite{Ter-Mikaelian1972}),
the emission spectrum in a randomly oriented crystal contains a coherent part
in addition to the Bethe-Heitler term which characterized which characterized
the incoherent bremsstrahlung in an amorphous medium.
Therefore, some discrepancy, seen in the inset, between the calculated dependencies and
the Bethe-Heitler values can be attributed to the contribution of the coherent term.}

By normalizing the channeling spectral intensities to the Bethe-Heitler values,
the enhancement spectral factors can be obtained for the channeling radiation by
the positrons and electrons.
These factors were computed using two sets of the simulated trajectories for each of
the projectiles.
The first set, discussed above, corresponds to the case when
the velocity of a projectile at the crystal entrance is parallel to Si(110)
plane, i.e., the incident angle $\psi$ is zero.
The second set of the trajectories was simulated allowing the incident angle
to be uniformly distributed within the interval $[-\psi_{\rm L}, \psi_{\rm L}]$
with $\psi_{\rm L}=62$ $\mu$rad being Lindhard's planar critical value
calculated in accordance with Eq. (1) from
the paper \cite{BakEtal1985}.

\begin{figure}[!t]
\centering
\includegraphics[width=\textwidth,keepaspectratio=true]{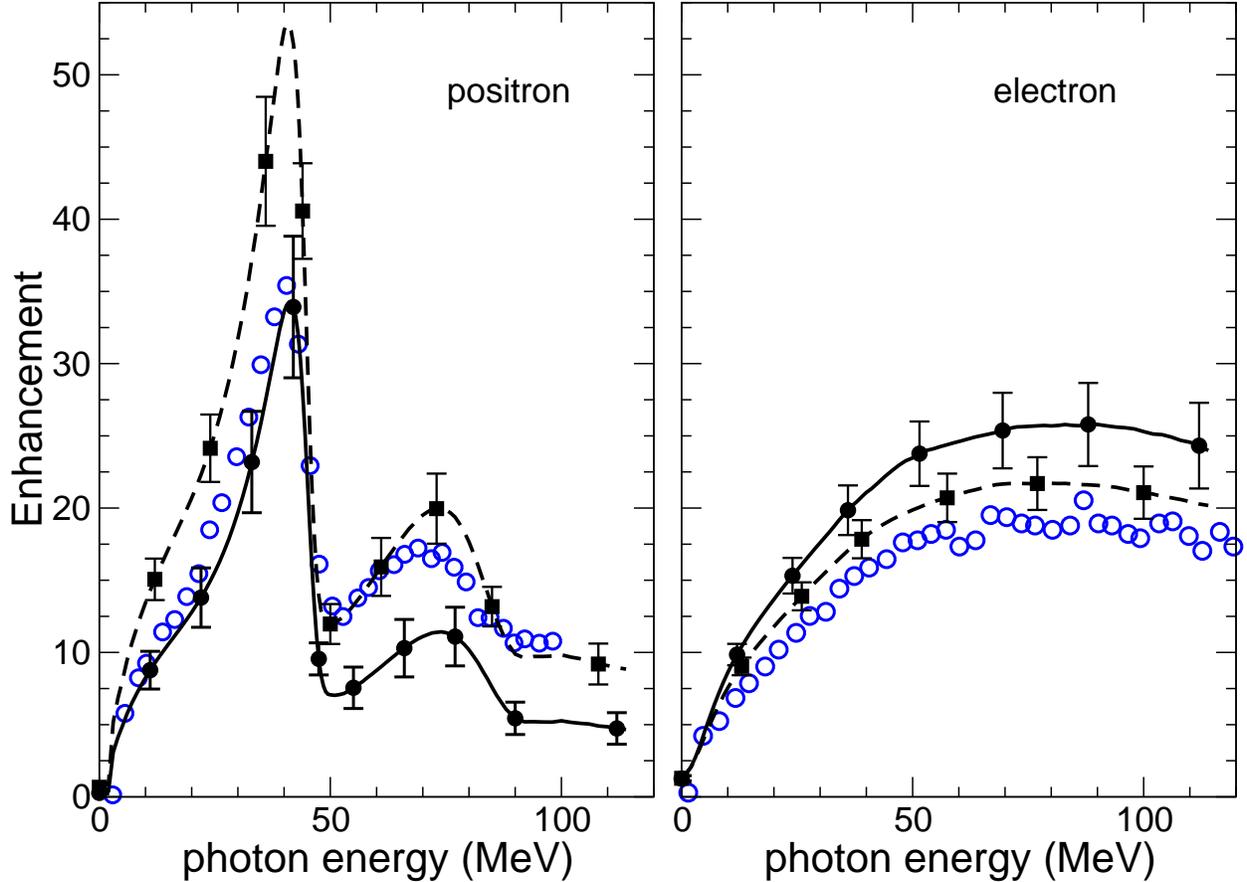}
\caption{Enhancement factor of the channeling radiation over the Bethe-Heitler
spectrum. The left and right plots are for the positrons and electrons, respectively.
Open circles stand for the experimental data \cite{BakEtal1985}.
Solid curves correspond to the calculations shown in
Fig.~\ref{dE_6_7GeV_el_pos_v04.fig}
and correspond to the zero incident angle, $\psi=0$.
Dashed curves correspond to the calculations with the incident angle lying within
$\psi=[-\psi_{\rm L}, \psi_{\rm L}]$ with $\psi_{\rm L}=62$ $\mu$rad
{(s}ee also explanation in the text{)}.
{
The curves are drawn over ca. 200 photon energy points in which the spectra
were calculated.
The symbols mark a small fraction of the points
and are drawn to illustrate typical statistical errors
(due to a finite number of the simulated
trajectories) in different parts of the spectrum.}
}
\label{Enhancement_6_7GeV_el_pos_035mrad_v03a.fig}
\end{figure}

The calculated enhancement factors are compared in
Fig.~\ref{Enhancement_6_7GeV_el_pos_035mrad_v03a.fig} with the experimental
results \cite{BakEtal1985} for 6.7 GeV projectiles \footnote{In the main text
of the cited paper these data refer to 7 GeV projectiles. However,
in the Note added in proof it is indicated that the actual beam momentum
is 4 \% lower.
This is also stressed in the paper \protect\cite{Uggerhoj1993} where
the beam energy of 6.7 GeV, used at the experiments, is indicated.}.
The open circles stand for the experimental data obtained by digitizing
Fig.~12 from the cited paper.
The solid and dashed curves represent the calculated dependencies for the
two sets of trajectories as indicated in the caption.
Fig.~\ref{Enhancement_6_7GeV_el_pos_035mrad_v03a.fig} demonstrates
that the
results of simulation reproduce rather well the shape of the
spectra and, in the case of the positron channeling, the positions of the main and
the secondary peaks.
With respect to the absolute values both calculated spectra,  $\psi=0$
and $|\psi|\leq \psi_{\rm L}$, exhibit some deviations from the
measured dependencies.

For positrons, the curve with $\psi=0$ perfectly matches the experimental data
in vicinity of the main peak but underestimates the measured yield of the higher
harmonics.
Increasing the incident angle results in some overestimation of the main
maximum but improves the agreement above $\hbar\om=40$ MeV.
For electrons, the $\psi=0$ curve exceeds the measured values, however,
the increase in $\psi$ leads to a very good agreement if one takes into
account the statistical errors of the calculated dependence.

The aforementioned deviations can be due to several reasons.
Modeling a crystalline field as a superposition of the
atomic fields described by the Moli\`{e}re potentials
can lead to intrinsic errors.
Though the Moli\`{e}re approximation is a well established and efficient
approach, more  realistic schemes for the crystalline fields, based, for example,
on X-ray scattering factors \cite{DoyleTurner,ChouffaniUberall1999},
can also be employed for the channeling simulations.

Another source of the discrepancies can be attributed to some uncertainties
in the experimental set-up described elsewhere \cite{BakEtal1985,Uggerhoj1993}.
In particular, it was indicated that the incident angles were in the interval
$[-\psi_{\rm L}, \psi_{\rm L}]$ with the value $\psi_{\rm L}=62$ $\mu$rad
for a 6.7 GeV projectile.
However, no clear details were provided on the beam emittance which becomes an
important factor for comparing theory vs experiment.
In our calculations we used a uniform distribution of the particles within
the indicated interval of $\psi$, and this is also a source of the uncertainties.
We have also simulated the spectra for larger cutoff angle equal to $2\psi_{\rm L}$
(these curves are not presented in the figure).
It resulted in a considerable ($\approx 30$ \%) decrease of
the positron spectrum in the vicinity of the first harmonic peak.
However, a rigorous simulation of the emittance properties requires additional
substantial computational efforts and we reserve its implementation for
future studies.

On the basis of the comparison with the experimental data we conclude that
our code produces reliable results and can be further used to simulate the propagation 
of ultra-relativistic projectiles along with the emitted radiation.

\subsection{Results for 855 MeV electrons and positrons}
\label{855MeV_Si}

Another relevant benchmark for our simulations are the channeling properties
of 855 MeV electrons and positrons in Si(110) that have been addressed
in previous theoretical
\cite{KKSG_simulation_straight,KKSG_NuovoCimento} %
and experimental
\cite{Backe_EtAl_2008,Backe_EtAl_2010a,Backe_EtAl_2010,Backe_EtAl_2011,%
Backe_EtAl_2011a} studies.
To this end, we have performed extensive calculations of the particles
trajectories and the emitted radiation spectra formed in $L=50$ $\mu$m and
$150$ $\mu$m crystalline and amorphous silicon.

For propagation along the Si(110) crystallographic plane, qualitative
character of the electron and positron trajectories was observed to be the
same as for 6.7 GeV projectiles discussed in the previous section.
For this reason we do not present the illustrative figure with the
simulated trajectories.
The dechanneling length of a 855 MeV positron, estimated as described above,
is $\Ld\approx 570$ $\mu$m.
Therefore, most of the positrons traverse the crystals in the channeling mode.
However, the considered crystal lengths were large enough to  deduce quantitative
information on the electron dechanneling lengths and to compare the result
with the previous studies \cite{KKSG_simulation_straight,Backe_EtAl_2008}.

To determine the electron dechanneling length each simulated trajectory
(of a total number $\approx 3000$) was analyzed with respect to comprising
segments of the channeling motion.
The particle was considered to be in the channeling mode if
it crossed the channel mid-plane at least three times, i.e. completed
one full oscillation between the channel boundaries.
Not all the particles become captured into the channeling mode
at the crystal entrance.
The fraction $\calA$ of the accepted electrons was found to be $\approx 0.65$
of the total number of the incident particles.
For the accepted particles the following two penetration depths
$\Lp$ were calculated.
The first one, $L_{\rm p1}=11.69 \pm 0.64$ $\mu$m was found as a mean value
of the primary channeling segments, which started at the entrance and lasted till
the dechanneling point somewhere inside the crystal.
Generally speaking, this quantity is dependent on the angular distribution
of the particles at the entrance.
The cited value of $L_{\rm p1}$ was obtained for a zero-emittance beam
collimated initially along the (110) planar direction.
Thus, it was meaningful to calculate another penetration depth, $L_{\rm p2}$,
defined as  a mean value of all channeling segments, including those which
appear due to the rechanneling.
In the rechanneling process an electron is captured into the channeling mode
having, statistically, an arbitrary value of the incident angle $\psi$
not greater than  Lindhard's critical angle.
Therefore, $L_{\rm p2}$ mimics the penetration depth of the beam with
a non-zero emittance $\approx \psi_{\rm L}$.
The calculated value $L_{\rm p2}= 10.9 \pm 0.3$ $\mu$m turned out to be not
much smaller than $L_{\rm p1}$, especially taking into account statistical
uncertainties.
The decrease of the confidence interval for $L_{\rm p2}$ is related to the
increase in the number of the channeling events
(approximately by a factor of 3.5) due to the rechanneling.

Either of the calculated quantities $L_{\rm p1,2}$ can be used as an
{\em estimate} of the dechanneling length.
In this connection, it is worth noting that the cited values
are noticeably larger than the dechanneling length $8.26 \pm 0.08$ $\mu$m
calculated earlier \cite{KKSG_simulation_straight}.
This difference can be attributed to a peculiar model used in
the cited paper to describe the electron--atom scattering.
In {Appendix}~\ref{Snapshot}
we demonstrate that the model overestimates the scattering angle,
leading, thus, to a decrease in the dechanneling length.
On the other hand, the presented values of $L_{\rm p1,2}$
are smaller than $\Ld=18$ $\mu$m, obtained  \cite{Backe_EtAl_2008}
within the framework of the diffusion theory.
The nature of this discrepancy is still to be understood.

\begin{figure}[!t]
\centering
\includegraphics[width=\textwidth,keepaspectratio=true]{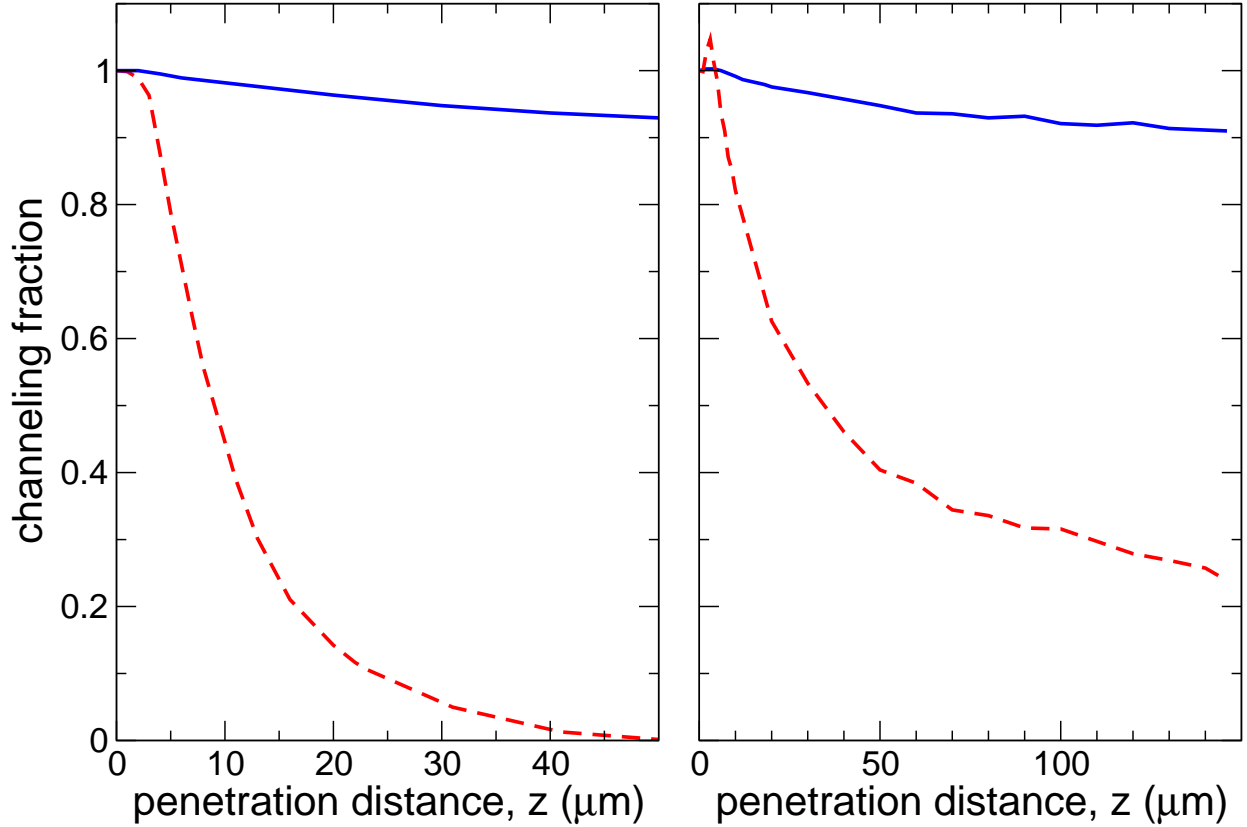}
\caption{Channeling fractions for 855 MeV positrons (solid lines) and electrons
(dashed lines) in Si(110) as functions of penetration depth.
The left panel corresponds to the particles remaining in the
{\em same channel} starting the entrance point $z=0$.
The right panel presents the dependence on $z$ of the total number of particles
moving in the channeling mode in {\em any channel}.
The fractions are determined with respect to the numbers of
initially accepted particles.
{The increase of the channeling fraction for electrons at small $z$, seen
on the right panel,
is due to the rechanneling effect (see also explanation in the text).}
}
\label{fractions.fig}
\end{figure}

For the sake of completeness, let us mention the results of
similar analysis carried out for  855 MeV positron channeling in $L= 150$ $\mu$m
Si(110) crystal.
In this case, the acceptance $\calA= 0.98$ is noticeably higher
due to the repulsive character of positron--atom interaction force,
which steers the projectile away from the nuclei and decreases the rate of hard
collision events.
As mentioned above, most of the positrons channel through the whole crystal.
Therefore, it is meaningful to calculate the penetration length due to
the primary channeling events only.
The obtained value $L_{\rm p1} = 133.8 \pm 2.7$ $\mu$m
can be considered only as a lower bound of the positron dechanneling length.

To quantify the channeling properties, we have also computed
fractions of the channeling particles versus penetration distance $z$.
Two types of the fractions were considered:
(i) for the particles remaining in the {\em same channel} where to they
were captured at the entrance;
(ii) for the particles which become trapped into {\em any channel} in
the course of propagation due to the rechanneling process.
Both fractions were determined with respect to the numbers of the
particles accepted at
the entrance, and, thereby, the dependencies start from the value of one at $z=0$.

The results of calculations are shown in Fig.~\ref{fractions.fig}.
The fractions of channeling positrons decrease very slow with $z$.
In the case of electrons the decay is much more rapid.
For example,  half of the primarily channeled electrons propagate
till the distance $z\approx 9.14$ $\mu$m,
and practically none of them channel up to $z=50$ $\mu$m
(see the left panel).
The fractions with account for the rechanneling are shown in the right panel
of the figure.
In the case of positron channeling the rechanneling events are very rare,
and, therefore, there is no visible change in comparison with the behaviour of
the primarily channeled fraction.
For electrons, on the contrary, the exponential decay is substituted with a
much slower one.
This effect has been noted earlier\cite{KKSG_simulation_straight} and it was
shown that the fraction of the channeling particle with account for the
rechanneling decreases as $\propto z^{-1/2}$.
To be noted is an increase of the channeling fraction at small penetration depths,
which is due to electrons captured into the channeling mode right after the
entrance point.

\begin{figure}[!t]
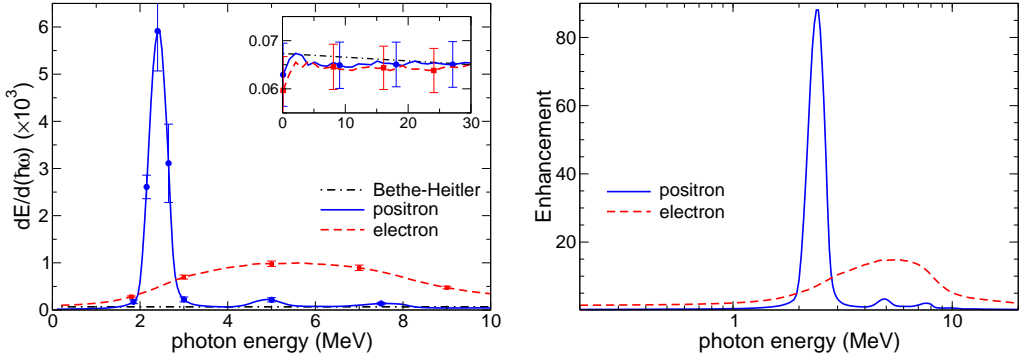

\centering
\includegraphics[scale=0.27,clip]{Figure_08left.eps}%
\hspace*{0.3cm}
\includegraphics[scale=0.27,clip]{Figure_08right.eps}
\caption{{\em Left plot}: Spectra for 855~MeV positrons (solid line) and
electrons (dashed line) passing through a 50~$\mu$m thick Si(110) crystal.
The inset shows the
spectra produced in amorphous Si and the Bethe-Heitler spectrum (dotted line).
{\em Right plot}: Spectral enhancement of the channeling radiation with
respect to the Bethe-Heitler values.}
\label{spectra_855MeV.fig}
\end{figure}

The electron and positron emission spectra calculated from the
simulated trajectories in $L=50$ $\mu$m straight Si(110) crystals
are shown on the left panel of Fig.~\ref{spectra_855MeV.fig}.
The spectra refer to the maximum emission angle $\theta_0= 21$ mrad
in order to provide the benchmark results for the experimental conditions
at MAMI setup \cite{BackeLauth_2012}.
To test the accuracy of our approach the spectra in amorphous Si were also
computed and are presented in the inset being compared the Bethe-Heitler spectrum
of the elastic bremsstrahlung calculated using Eqs. (\ref{BH:eq.06})
and (\ref{BH:eq.09}).
Good agreement between the spectra produced in the amorphous medium
allows us to conclude that the channeling
spectral simulations deliver reliable results.

On the right panel of Fig.~\ref{spectra_855MeV.fig}, we present the enhancements
of the radiation by positrons and electrons in Si(110) with respect to the
Bethe-Heitler calibration.
We hope the theoretical curves in this plot to be useful for
the ongoing experimental studies \cite{BackeLauth_2012}.

\section{Conclusion and Outlook}
\label{Conclusion}

We have described the newly developed 
code, which was
implemented in the \MBNExplorer package \cite{MBN_ExplorerPaper,MBN_ExplorerSite}
to simulate trajectories of an ultra-relativistic projectile in
a crystalline medium.
The description of the particle motion is given in classical terms by
solving the relativistic equations of motion which account for
the interaction between the projectile and the crystal atoms.
The probabilistic element is introduced into the scheme by a random choice of
transverse coordinates and velocities of the projectile at the crystal
entrance as well as by accounting for the random positions of the crystal
atoms due to thermal vibrations.
The simulated trajectories can be used as the input data for
numerical analysis of spectral-angular and spectral distributions of the
emitted electromagnetic radiation.
The current version of the channeling module of \MBNExplorer and
supplementary documentation
are available to users at the \MBNExplorer website \cite{MBN_ExplorerSite}
upon registration and agreeing with the license conditions.

Being a part of \MBNExplorer the 
code takes advantage of particular
algorithms, implemented in the package, which facilitate numerical procedures.
Additionally, the applicability of the code to different crystalline structures
can be adjusted either by choosing a proper interaction potential from a large
variety of the potentials already included in the package or, if necessary,
by including a new potential.
We would like to stress that with minor modifications the code can be
generalized further allowing one to simulate the propagation of
charged relativistic projectiles in various media, such as
heterocrystalline structures (including superlattices),
bent and periodically bent crystals,
amorphous solids, liquids, nanotubes, fullerites,
biological environment, and many more.
In this connection it is worth noting that \MBNExplorer allows one
to optimize the structure of a complex molecular system.
For example, it can be used to optimize the geometry of strained-layer
Si$_{1-x}$Ge$_{x}$ superlattices  produced in the molecular beam epitaxy
laboratory of University of Aarhus and used in the channeling experiments at
the Mainz Microtron MAMI  \cite{Backe_EtAl_2010,Backe_EtAl_2011}.
Therefore, by means of the \MBNExplorer package it is possible to
generate a realistic structure of the medium {\em and} to carry out the
simulations of the trajectories.

Two case studies have been carried out for initial approbation and
verification of the code. 
Differing in the projectile energy, $\E=6.7$ GeV and $\E=855$ MeV,
both case studies refer to the simulation of trajectories and
calculation of spectral distribution of the radiation emitted by
ultra-relativistic electrons and positrons moving in oriented
Si(110) crystal and in amorphous silicon.

For the incident energy $6.7$ GeV the calculated spectra were compared
with the available experimental data for Si(110) \cite{BakEtal1985,Uggerhoj1993}
and with predictions of the Bethe-Heitler theory
for the amorphous environment.
For both projectiles a good agreement has been found between
theory and experiment.
Coincidence of the simulated radiative spectra of electrons and positrons
with each other and with the Bethe-Heitler results provide additional
indication of the reliability of the code. 

The case study of an $\E=855$ MeV light projectile channeling in Si (110)
is of interest in connection with  the ongoing experiments with electron beams
at Mainz Microtron \cite{Backe_EtAl_2011}
and possible experiments with the positron beam \cite{Backe_EtAl_2011a}.
By analyzing the simulated trajectories we estimated the electron
dechanneling length $\Ld$ independent on the angular distribution of the beam
particles at the entrance, as a mean value of all channeling segments.
The obtained result, $10.9 \pm 0.3$ $\mu$m, exceeds by approximately 30
per cent the value calculated recently \cite{KKSG_simulation_straight}.
Apart from some difference in the definitions of the dechanneling length,
this discrepancy can be attributed to a specific model used in
{earlier studies} \cite{KKSG_simulation_straight}
to describe electron--atom elastic scattering.
We have demonstrated that the model overestimates the mean scattering angle,
and, thus, underestimates the dechanneling length.
On the other hand, our estimate for $\Ld$ is lower than the
reported experimental value \cite{Backe_EtAl_2008}.
To clarify this discrepancy it will be instructive to compare
the calculated enhancement factor of the channeling radiation over
the incoherent bremsstrahlung background with the corresponding experimental
data once it becomes available.

As a prime further step in application of the developed code,
the simulation of trajectories  and
calculations of spectral intensities of the radiation emitted
in crystalline undulators are to be performed and compared with the
experimental results available for electrons of various energies
\cite{Backe_EtAl_2011}.
For the sake of comparison, the computations will be also carried out for
positrons.
The results of this work, which is currently in progress, will be published
elsewhere.

We also plan to introduce several new features to the numerical algorithm
described in section \ref{Algorithm}
aiming to expand the range of applicability of both the code for the trajectories
simulations and the one for the spectrum
calculation.
In particular, the first equation of motion (\ref{MC_Simulations.01})
will be supplemented with the radiative damping force which allows us to
account for radiation energy losses of light projectiles in tens to hundreds
GeV energy range.
Calculations of characteristics of the emitted radiation will be improved
by including a correction due to the density effect and by taking into
account the contribution of the transition radiation, formed
at the crystal entrance, to the total emission spectrum.

{\color{black}
\begin{acknowledgments}
In part, this work was supported by the European Commission (the IRSES-CUTE
project).
The code development was carried out by Gennady B. Sushko, Ilia A. Solov'yov, 
Andrei V. Korol and, Andrey V. Solov'yov  under the auspices
of Virtual Institute on Nano Films (VINF).
The possibility to perform complex computer simulations at the Frankfurt Center for
Scientific Computing is gratefully acknowledged.
Ilia A. Solov'yov acknowledges support as a Beckman Fellow.

\end{acknowledgments}
}

\appendix

\section{Classical Scattering of an Ultra-Relativistic Projectile from
a ``Snapshot'' Atom \label{Snapshot}}

The code for the simulation of the channeling of
ultra-relativistic charged projectiles, described in Refs.
\cite{KKSG_simulation_straight,KKSG_NuovoCimento}, was based
on the peculiar model of the elastic scattering of the projectile
from the crystal constituents.
The model assumes that due to the high speed of the projectile, its
interaction interval with a crystal atom is short enough
to substitute the atom with its ``snapshot'' image:
instead of the continuously distributed electron charge
the atomic electrons are treated as point-like charges placed at fixed
positions around the nucleus \footnote{We term
such a system as  ``a snapshot atom''.}.
Next, the model implies that the interaction of an ultra-relativistic
projectile with each atomic constituent can be reduced
to the classical Rutherford scattering.
Scattering events happen sequentially as the projectile flies by an atom.
The projectile trajectory is modeled by a piecewise linear curve the
vertices of which correspond to the events.
Between two successive events the projectile moves with a constant velocity $\bfv$.
The change of the transverse momentum $\Delta \bfp_{\perp}$ in the event
is calculated within the small scattering angle approximation, i.e.
as the integral of the impulse $\bfF_{\perp} \d t$ along the straight
line aligned with $\bfv$ (see, e.g., \cite{Landau1}).
As a result, the total scattering angle $\btheta$ acquired by the
projectile of the charge $\Zp e$ in the collision with a ``snapshot''
atom can be written in the following vector form:
\begin{eqnarray}
\btheta_{\rm S} (\{\bfr_j\})
=
\sum_{j=1}^{Z+1}\btheta_j
\approx
{2\Zp e\over \E}
\sum_{j=1}^{Z+1} q_j\,{\brho_j \over \rho_j^2}
\,,
\label{SnapshotTheta.01}
\end{eqnarray}
where the subscript ``S'' stands for the ``snapshot'' atom,
$\{\bfr_j\}\equiv \bfr_1,\bfr_2, \dots, \bfr_Z$
are the position vectors of $Z$ atomic electrons.
The sum is carried out over the atomic constituents:
the nucleus (the charge $q_j=Ze$) and the electrons ($q_j=-e$).
For each constituent the index $j$ equals to the ordering number of the
event  in the sequence of all $Z+1$ scattering events.
In the small-angle approximation the scattering angle in the $j$th
collision with a point-like charge $q_j$ is calculated as
$\btheta_j\approx 2\Zp e q_j\,\brho_j /\E\rho_j^2$, where
$\rho_j$ is the impact parameter and $\brho_j$ is perpendicular to the
projectile velocity $\bfv$ before the collision.

As it was noted in Ref. \cite{KKSG_simulation_straight}, the above
procedure is approximate in a sense that it is restricted to the limit of
small scattering angles when $|\btheta_{j}|\ll 1$ and
$|\btheta_{\rm S}|\ll 1$.
It was stated, that in the opposite limit not only Eq. (\ref{SnapshotTheta.01})
is not valid but also the ``snapshot atom'' concept is wrong.
However, the large angle scattering is not important for modeling
the channeling process.
Therefore, one can rely on the described procedure provided it is valid
for the scattering angles smaller that Lindhard's
critical angle $\phi_{\rm L}$ which is typically in the submilliradian
range for ultra-relativistic projectiles.

In what follows we demonstrate, that despite seeming credibility of the
``snapshot'' model it noticeably overestimates the mean scattering angle
in the process of elastic scattering.
Qualitatively, it is clear that substituting a ``soft'' electron cloud
with a set of point-like static electrons must lead to the increase of
the scattering angle simply because each electron acts as a charged
scatterer of an infinite mass.
As a result, the projectile experiences, on average, harder collisions
with electrons as compared to the case when they are continuously
distributed in the space.

To illustrate the above statement let us calculate the root-mean-square
(r.m.s.) scattering angle
$\overline{\theta}_{\rm S}(b) = \sqrt{\langle \btheta_{\rm S}^2\rangle}$
as a function of the initial impact parameter $b$ with respect to the
nucleus and, then, compare the result with the dependence $\theta_{\rm M}(b)$ 
obtained for the atom treated in the Moli\`{e}re approximation.

The mean square scattering angle $\langle \btheta_{\rm S}^2\rangle$
is calculated by averaging the square of the right-hand side of Eq.
(\ref{SnapshotTheta.01}) over the ensemble of the ``snapshot'' atoms:
\begin{eqnarray}
\langle \btheta_{\rm S}^2\rangle
=
{1\over N} \sum_{a=1}^{N}
\btheta_{\rm S}^2 (\{\bfr_j\}_a)
\approx
\left({2\Zp e\over \E}\right)^2
{1\over N} \sum_{a=1}^{N}
\left(\sum_{j=1}^{Z+1} q_{aj}\,{\brho_{aj} \over \rho_{aj}^2}\right)^2
\label{SnapshotTheta.08}
\end{eqnarray}
where the subscript $a=1,2,\dots, N$ enumerates the atoms.

To construct a ``snapshot'' atom one has to randomly generate
the position vectors $\{\bfr_j\}_a\equiv \bfr_{a1},\bfr_{a2},\dots,\bfr_{aZ}$
of its electrons.
For doing this we follow the scheme described in Ref.
\cite{KKSG_simulation_straight}.
The scheme implies spherical symmetric distribution of the direction
of the position vectors whereas the distance $r_j$ from the
nucleus for each atomic electron is found by solving the equation:
\begin{eqnarray}
\chi(r_j) - r_j\chi^{\prime}(r_j) = \xi_j\,,
\label{Snapshot.40}
\end{eqnarray}
where $\chi(r)$ stands for the Moli\`{e}re screening function defined in
(\ref{MC_Simulations.03})
and $\xi_j$ is a uniform random deviate between $0$ and $1$.
In the current work, we used the routine {\tt ran2} from Ref. \cite{NumRec}
to generate $\xi_j$.

It is noted in Ref. \cite{KKSG_simulation_straight} that if the positions
of the electrons are chosen as described above, then the electrostatic
potential $U_{\rm S}$ of an  atom averaged over the ensemble of
the ``snapshot'' atoms reproduces the Moli\`{e}re potential:
\begin{eqnarray}
\left\langle U_{\rm S} \right\rangle(r)
=
\lim_{ N\to \infty}
{1\over N} \sum_{a=1}^{N}
\left(
{Z e\over r}
-
e
\sum_{j=1}^{Z}
 {1 \over |\bfr - \bfr_{aj}|}
\right)
=
U_{\rm M}(r)\,.
\label{UandE_snapshot.01}
\end{eqnarray}
\begin{figure}[!t]
\centering
\includegraphics[width=13cm,clip]{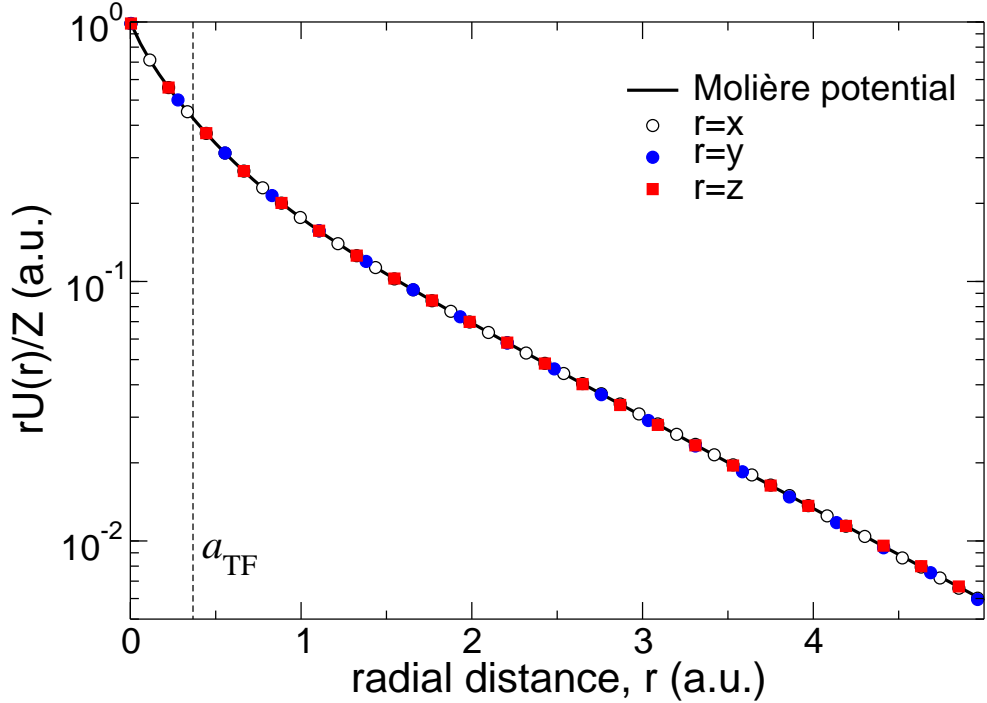}
\caption{
The ratio $rU(r)/Z$ versus radial distance calculated for a silicon
atom ($Z=14$).
The solid curve corresponds to the electrostatic potential $U(r)$
within the Moli\`{e}re approximation, Eq. (\ref{MC_Simulations.03}).
The dots represent the dependencies obtained for the averaged
potential $\left\langle U_{\rm S} \right\rangle(r)$, Eq.
(\ref{UandE_snapshot.01}), with the radial distance $r$ measured along
the $x,y,z$ directions, as indicated.
The averaging was carried out for $N=750000$  ``snapshot'' atoms.
The vertical line marks the Thomas-Fermi radius of a Si atom:
$\aTF = 0.194$ \AA $= 0.367$ a.u.
}
\label{Moliere_vs_MC_U_v02.fig}
\end{figure}
Figure (\ref{Moliere_vs_MC_U_v02.fig}) compares the dependencies
$rU(r)/Z$ calculated for a silicon atom within the Moli\`{e}re
approximation (solid line) and by means of the averaging procedure
(\ref{UandE_snapshot.01}) with $N=750000$.
The circles correspond to the potential
$\left\langle U_{\rm S} \right\rangle(r)$ calculated for the radial distances
measured along three different spatial directions.
The deviation of the simulated dependencies
$\left\langle U_{\rm S} \right\rangle(r)$ from the potential
$U_{\rm M}(r)$ is on the level of 0.1 \% for $r \leq 2$ a.u.$\approx 6\aTF$
and increases up to 1 \% for $r\gtrsim 5$ a.u.
Hence, we state that the averaging procedure (\ref{UandE_snapshot.01})
accompanied with the random generation of radial distances
(\ref{Snapshot.40}) reproduces the  Moli\`{e}re potential quite accurately.

Despite the agreement in the electrostatic potential evaluation,
the mean scattering angle calculated within the ``snapshot'' model
noticeably exceeds the scattering angle $\theta_{\rm M}$ of an
ultra-relativistic projectile in collision with the Moli\`{e}re atom.
In the small-angle  limit \cite{Landau1} one derives the following
dependence of $\theta_{\rm M}$ on the impact parameter $b$
for projectile electron or positron ($|\Zp|=1$):
\begin{eqnarray}
\theta_{\rm M}(b)
&\approx&
{1\over \E}
\left|\int_{-\infty}^{\infty} F_{\perp}\, \d z \right|
=
{2 e^2 Z\, b\over \E}
\left.
\int_{0}^{\infty}
{\chi(r) -r\chi^{\prime}(r) \over r^3}\, \d z\right|_{r=\sqrt{b^2 + z^2}}
\nonumber\\
&=&
{2 e^2 \over \E}\,
{Z \over \aTF}
\sum_{j=1}^3 \alpha_j\beta_j
K_1\left(\beta_j \,{b \over \aTF}\right)
\label{MoliereTheta.11}
\end{eqnarray}
where $K_1(\zeta)$ stands for the MacDonald function of the first
order (see, e.g., \cite{Gradshteyn}).
For small impact parameters, $b \ll \aTF$, one utilizes the
relation $K_1(\zeta\ll 1) \approx \zeta^{-1}$ and
derives
$\theta_{\rm M}(b \ll \aTF) = {2 Z e^2 / \E \, b }$
which is the scattering angle in the point Coulomb field of
the charge $Ze$.
In the limit of large argument
the MacDonald function behaves as $K_1(\zeta) \propto \xi^{-1/2}\exp(-\xi)$.
Therefore, the scattering angle $\theta_{\rm M}(b)$ decreases
exponentially for $b \gg \aTF$.

\begin{figure}[!t]
\centering
\includegraphics[width=13cm,clip]{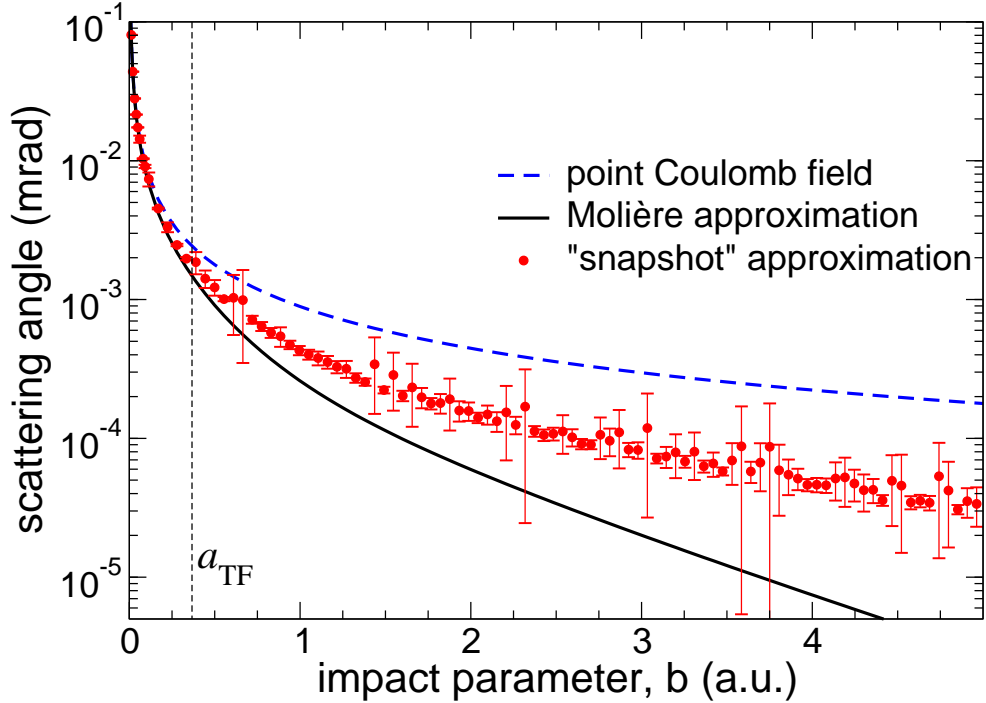}
\caption{
Scattering angle as a function of the impact parameter for a 855 MeV
electron scattered from a silicon atom ($Z=14$).
The solid curve corresponds dependence within the Moli\`{e}re
approximation, the dashed one -- to that
in the Coulomb field of the bare nucleus.
The filled circles with error bars stand for the
r.m.s. $\overline{\theta}_{\rm S}(b) =
\sqrt{\langle \btheta_{\rm S}^2\rangle}$ calculated in the ``snapshot''
approximation, Eq. (\ref{SnapshotTheta.08}).
The averaging was carried out for $N=500000$  ``snapshot'' atoms.
The vertical line marks the Thomas-Fermi radius of a Si atom:
$\aTF = 0.194$ \AA $= 0.367$ a.u.
}
\label{Coul_Mol_MC_allTheta.fig}
\end{figure}

In Fig. \ref{Coul_Mol_MC_allTheta.fig} we present the scattering angle
vs the impact parameter for the collision of a 855 MeV electron with
a silicon atom.
The calculations were performed within the Moli\`{e}re approximation,
Eq. (\ref{MoliereTheta.11}), and for a ``snapshot'' atom
and by means of the averaging procedure (\ref{SnapshotTheta.08})
with $N=500000$.
For the sake of comparison, the dependence
$\theta(b) = {2 Z e^2 / \E \, b }$, which characterizes the
scattering angle in the Coulomb field of the bare nucleus, is also plotted.

For small impact parameters $b\ll \aTF$, where the collisional process
is mainly governed by the interaction with the nucleus, all three approaches
lead to the same dependence $\theta \propto 1/b$.

The deviation of the $\overline{\theta}_{\rm S}(b)$ curve from
the Moli\`{e}re one becomes clearly visible $b \gtrsim \aTF$ and
is steadily more pronounced with further increase of the impact
parameter.
For $b \gg \aTF$ the slope of $\overline{\theta}_{\rm S}(b)$ is more
point-Coulomb like rather than an exponential decrease exhibited
by the function $\theta_{\rm M}(b)$ obtained for a short-range potential.
Additionally, one can see random jumps in the
$\overline{\theta}_{\rm S}(b)$ dependence accompanied with large error bars.
Both of these features, -- the Coulomb-like slope and
random irregularities, can be easily understood.
Indeed, as it was already pointed out, the main drawback of the ``snapshot''
model is in the assumption that not only the nucleus but also
{\em all} atomic electrons  are treated as motionless (and, thus, infinitely
heavy) point charges.
Therefore, the recoil of the scatterer is fully ignored in the collisional
process of a projectile with any of the atomic constituents.
Physically, it means that, on average, the collisions become ``harder''
so that the r.m.s. scattering angle increases.
Simultaneously, random positioning of electrons in a ``snapshot'' atom
may result very hard scattering events even for the distances $b \gg \aTF$.
These events, although being rare, lead to visible jumps and large
uncertainties in the $\overline{\theta}_{\rm S}(b)$ values.

Despite the fact that the absolute values of the scattering angle for
$b>\aTF$ are comparatively small, the deviation of
$\overline{\theta}_{\rm S}(b)$ from $\theta_{\rm M}(b)$ in this domain
influences the mean square angle $\langle \Theta^2 \rangle$
for a single scattering.
The latter quantity is very important in the multiple-scattering region,
where there is a large succession of small-angle deflections symmetrically
distributed about the incident direction.
The quantity $\langle \Theta^2 \rangle$ is proportional to the
following integral (see, e.g., \cite{Jackson}):
$\langle \Theta^2 \rangle \propto \int \theta^2\,
{\d\sigma \over \d \Om} \,\d \Om$
where $\d\sigma /\d \Om$ stands for the cross section of elastic scattering
differential with respect to the scattering angle.
Within the framework of classical mechanics
and in the limit of small scattering angles $\theta \ll 1$
the cross section reads (see, e.g., \cite{Landau1}):
\begin{eqnarray}
{\d\sigma \over \d \Om}
\approx
\left|{\d b \over \d \theta}\right|
{b(\theta) \over \theta}\,.
\label{CrossSection.01}
\end{eqnarray}
Using (\ref{CrossSection.01}) in the definition of
$\langle \Theta^2 \rangle$
one derives the following expression for the
ratio of the mean square angles obtained within the ``snapshot'' and the
Moli\`{e}re approximations:
\begin{eqnarray}
{ \langle \Theta^2 \rangle_{\rm S} \over \langle \Theta^2 \rangle_{\rm M}}
\approx
\left.
\int\limits_{b_{\min}}^{\infty}
\langle \btheta_{\rm S}^2\rangle(b)\,  b\, \d b
\right/
\int\limits_{b_{\min}}^{\infty}
\theta_{\rm M}^2(b)\,   b\, \d b\,.
\label{CrossSection.04}
\end{eqnarray}
To carry out the integrals one has to introduce a particular value of the
cutoff impact parameter $b_{\min}$.
\textcolor{black}{
For deducing the latter we point out the ``snapshot'' approximation was
introduced in Ref. \cite{KKSG_simulation_straight} as a part of the
code aimed at the adequate description of the channeling process.
From this end, the model must adequately describe the scattering
process at the distances $b\gtrsim \aTF$ between the projectile
and the crystal plane.
Hence, it is instructive to use $\aTF$ as the cutoff.
Then, Eq. (\ref{CrossSection.04}) produces
$\langle \Theta^2 \rangle_{\rm S}/\langle \Theta^2 \rangle_{\rm M}\approx 2.5$.
Within the diffusion theory of the dechanneling process (see, e.g.,
\cite{BiryukovChesnokovKotovBook,Backe_EtAl_2008}) the mean square angle
due to soft collisions defines the diffusion coefficient which, in turn,
is proportional to the dechanneling length $\Ld$.
In this context, the fact that the ``snapshot'' model overestimates
$\langle \Theta^2 \rangle$ for $b\gtrsim \aTF$
explains the aforementioned discrepancy in the $\Ld$
values for 855 MeV electrons in Si (110) calculated in
Ref. \cite{KKSG_simulation_straight} and in the current work, see
Section \ref{855MeV_Si}.
}

\section{The Bethe-Heitler Approximation: Collection of Formulae}
\label{BH}

In the {elementary process} of bremsstrahlung (BrS) a charged projectile
emits a photon being accelerated by the static field of a target
(nucleus, ion, atom, etc).

For ultra-relativistic projectiles, the Bethe-Heitler (BH) approximation
\cite{BetheHeitler1934} (with various corrections due to Bethe \textit{et al.}
\cite{BetheMaximon1954,DavisBetheMaximon1954} and
Tsai \textit{et al.} \cite{Tsai1974}) is the simplest and the most widely used one.

For the sake of reference below in this Section we present
the relevant formulae for the case of ultra-relativistic electrons/positrons
scattering from a neutral atom treated within the Moli\`{e}re approximation
\cite{Moliere}.

Starting from Eq. (3.80) in Ref. \cite{Tsai1974},
one can write the following formula for
the cross section differential with respect to the photon energy $\hbar\om$
and to the emission angle $\Om=(\theta,\phi)$
(but integrated over the angles of the scattered electron):
\begin{eqnarray}
\nonumber
{\d^2 \sigma \over \d(\hbar \om) \d \Om}
&=&
{4\alpha r_0^2 \over \pi}\,
{\gamma^2 \over \hbar \om}
\left\{
\left(
2-2x+x^2
- {4 (1-x) \over 1+\xi}
+ {4 (1-x) \over (1+\xi)^2}
\right)
{\calF - 1 + \ln(1+\xi) \over (1+\xi)^2}
\right.
\\
\label{BH:eq.01}
&&
\left.
-
Z(Z+1)\left(
  1
- {6 \over 1+\xi}
+ {6 \over (1+\xi)^2}
\right)
{1-x\over (1+\xi)^2}
\right\} .
\end{eqnarray}
Here $\alpha\approx 1/137$ is the fine structure constant,
$r_0=e^2/mc^2\approx 2.818\times 10^{-13}$ cm is the classical electron radius,
$x = {\hbar \om / \E}$ and $\xi=(\gamma\theta)^2$.
The factor $\calF$ is defined by Eqs. (3.5), (3.44) and (3.45) from
 Ref. \cite{Tsai1974}.
In the ultra-relativistic limit (more exactly, for $\gamma \gtrsim 10^3$)
it can be written as follows:
\begin{eqnarray}
\calF
=
Z^2
\left(\ln{184 \over Z^{1/3}} -1 - f\Bigl(\alpha Z)^2\Bigr)\right)
+
Z\left(\ln {1194 \over Z^{2/3}} - 1\right)\,,
\label{BH:eq.02}
\end{eqnarray}
where the function
$f\Bigl((\alpha Z)^2\Bigr) = (\alpha Z)^2
\sum_{n=1}^{\infty} \left[n^2\left(n^2 + (\alpha Z)^2\right) \right]^{-1}$
(with $\zeta =\alpha Z$) is the Coulomb
correction to the first Born approximation
worked out in Refs. \cite{BetheMaximon1954,DavisBetheMaximon1954}.
In the limit $(\alpha Z)^2 \ll 1$ the term $f\Bigl((\alpha Z)^2\Bigr)$
can be ignored.
For example, for a Si atom ($Z=14$)
$f\left((\alpha Z)^2\right) \approx 0.0126 \ll 1$.

The term proportional to $Z^2$ on the right-hand sides of
(\ref{BH:eq.01}) and (\ref{BH:eq.02}) stands
for the contribution of the elastic BrS process in which
the target atom does not change its state during the collision.
The terms $\propto Z$ are due to the inelastic BrS channels,
when the atom becomes excited or ionized.

To calculate the cross section of BrS radiated into the cone with
the opening angle $\theta_0$ one integrates
Eq. (\ref{BH:eq.01}) over the emission angles
$\theta =[0, \theta_0]$ and $\phi = [0,2\pi]$.
The result reads
\begin{eqnarray}
\!\!\!\!\!\!\!\!\!\!\!\!\!\!\!\!
\left.{\d \sigma \over \d(\hbar \om)}\right|_{\theta\leq\theta_{0}}
&=&
{\d \sigma \over \d(\hbar \om)}
+
{4\alpha r_0^2 \over \hbar \om}
\left\{
Z(Z+1)
\left(
1 - {4\over D_0}
+ {26 \over 9D_0^2}
\right)
{1-x  \over D_0}
\right.
\nonumber\\
&-&
\left.\left(
2-2x+x^2
- {2 (1-x) \over D_0}
+ {4 (1-x) \over 3D_0^2}
\right)
{\calF + \ln D_0\over D_0}
\right\},
\label{BH:eq.03}
\end{eqnarray}
with $D_0 = 1+(\gamma\theta_0)^2$.
In the limit of large emission angles when $\theta_0 \gg 1/\gamma$
the second term on the right-hand side goes to zero.
Therefore, the first term stands for the
cross section differential in the photon energy but integrated over the whole
range of the emission angles.
Its explicit expression is as follows
(cf. Eq. (3.83) in Ref. \cite{Tsai1974}):
\begin{eqnarray}
{\d \sigma \over \d(\hbar \om)}
&=&
\int_0^{2\pi}
\d\phi
\int_{0}^{\infty}
\theta\, \d\theta
{\d^2 \sigma \over \d(\hbar \om) \d \Om}
\approx
{4\alpha r_0^2 \over 3\hbar \om}
\left(
(4 -4x + 3 x^2)\, \calF
+ Z(Z+1)\,{1-x\over 3}
\right)
\label{BH:eq.04}
\end{eqnarray}

To calculate the cross section of the elastic BrS
one substitutes $Z(Z+1) \to Z^2$ on the right-hand sides of
Eqs. (\ref{BH:eq.01}), (\ref{BH:eq.03}) and (\ref{BH:eq.04})
as well as ignores the last term in Eq. (\ref{BH:eq.02}).
The latter approximation leads to the following reduction:
\begin{eqnarray}
\calF
\to
\calF_{\rm el}
=
Z^2\left[\ln {184 \over Z^{1/3}} - 1 - f\left((\alpha Z)^2\right) \right]\,.
\label{BH:eq.05}
\end{eqnarray}
Then,
the single differential cross section of elastic BrS emitted within
the cone $0\leq \theta \leq \theta_0$ is given by:
\begin{eqnarray}
\!\!\!\!\!\!\!\!\!\!\!\!
\left.{\d \sigma_{\rm el} \over \d(\hbar \om)}\right|_{\theta\leq\theta_{0}}
&=&
{\d \sigma_{\rm el}\over \d(\hbar \om)}
+
{4\alpha r_0^2} \,{Z^2\over \hbar \om}
\left\{
{1-x  \over D_0}
\left( 1 - {4 \over D_0} + {26 \over 9D_0^2}
\right)
\right.
\nonumber\\
&-&
\left.
\left(
2-2x+x^2
- {2 (1-x) \over D_0}
+ {4 (1-x) \over 3D_0^2}
\right)
{\calF_{\rm el} + \ln D_0 \over Z^2 D_0}
\right\},
\label{BH:eq.06}
\end{eqnarray}
where
\begin{eqnarray}
{\d \sigma_{\rm el} \over \d(\hbar \om) }
=
{4\alpha r_0^2 \over 3}\,{Z^2\over \hbar \om}
\Bigl[
(4 -4x + 3 x^2)\, {\calF_{\rm el} \over Z^2}
+ {1-x\over 3}
\Bigr]
\label{BH:eq.07}
\end{eqnarray}
is the Bethe-Heitler spectrum of elastic BrS.

Within the framework of less accurate approximation,
used frequently for quantitative estimates (see, e.g.,
\cite{Akhiezer,BakEtal1985}),
one ignores the non-logarithmic terms in (\ref{BH:eq.05}):
\begin{eqnarray}
\calF_{\rm el}
\approx
\ln {184 \over Z^{1/3}}.
\label{BH:eq.08}
\end{eqnarray}

In order to calculate the spectral-angular distribution of the radiated energy
$\d^2 E / \d(\hbar\om)\d \Om$ in an amorphous target of the thickness $L$
 much less then the radiation length \cite{ParticleDataGroup2010}
one multiplies  Eq. (\ref{BH:eq.01}) by
the photon energy $\hbar\om$, by the volume density $n$ of the target atoms
and by $L$:
\begin{eqnarray}
{\d^2 E \over \d(\hbar \om)\d \Om}
=
n L\, \hbar \om\, {\d^2 \sigma\over \d(\hbar \om)\d \Om}\,.
\label{BH:eq.09}
\end{eqnarray}
Spectral distribution $\d E / \d(\hbar\om)\Bigl|_{\theta\leq\theta_{0}}$
of the energy radiated within the cone $\theta \leq \theta_0$ is
obtained from (\ref{BH:eq.09}) by
substituting the double differential cross section
either with
${\d \sigma_{\rm el} /\d(\hbar \om)}\Bigl|_{\theta\leq\theta_{0}}$
(for the total emitted energy)
or with
${\d \sigma_{\rm el} / \d(\hbar \om)}\Bigl|_{\theta\leq\theta_{0}}$
(if accounting for elastic BrS only).

\begin{figure}[!t]
\centering
\includegraphics[width=13cm,clip]{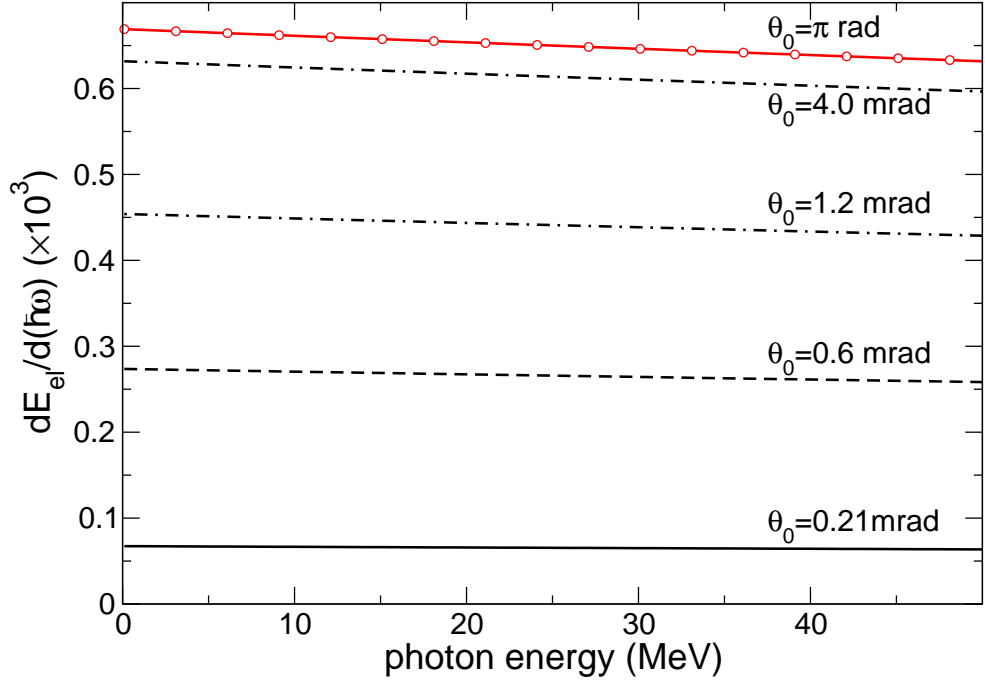}
\caption{
Bethe-Heitler spectra of the energy $\d E_{\rm el}/\d(\hbar\om)$
radiated via the elastic BrS channel by a $\E=855$ MeV electron
in amorphous silicon of the thickness $50$ $\mu$m.
Different curves correspond to different values of the
emission cone angle $\theta_0$ as indicated.
The curve $\theta_0=\pi$ rad stands for the spectral distribution
integrated over the whole range of the emission angles.
The elastic BrS cross section was calculated within the
logarithmic approximation, see Eq. (\ref{BH:eq.08}).
}
\label{BH_elastic_855MeV.fig}
\end{figure}

For illustrative purposes in Fig.~\ref{BH_elastic_855MeV.fig} we show the
spectral distributions $\d E_{\rm el} / \d(\hbar\om)\Bigl|_{\theta\leq\theta_{0}}$
of elastic BrS formed during the passage of a $\E=855$ MeV electron
through a 50 $\mu$m thick amorphous silicon ($n=5\times10^{22}$ cm$^{-3}$).
The curves were calculated for different values of the emission cone
angle as indicated.
The value $\theta_0=0.21$ mrad corresponds to the limit of small
emission angles $(\gamma\theta_0)^2 \ll 1$ where
$\gamma^{-1}\approx 6\times10^{-3}$ for the indicated incident energy.
For each photon energy the magnitude of
$\d E_{\rm el} / \d(\hbar\om)\Bigl|_{\theta\leq\theta_{0}}$
steadily increases with $\theta_0$ reaching its upper limit
at $\theta_0=\pi$ which corresponds to the cross section integrated over
the whole range of the emission angle, see Eq. (\ref{BH:eq.07}).

\bibliographystyle{apsrev}
\bibliography{Channeling_bib}
\end{document}